\documentclass[aps,prc,preprint,floatfix]{revtex4}

\usepackage{amsmath}
\usepackage{graphicx}
\usepackage{bm}
\usepackage{color}
\usepackage[colorlinks=true,linkcolor=blue]{hyperref}
\newcommand{\eps}{\varepsilon}

\preprint{JLAB-THY-13-1707, ADP-13-15/T835}

\begin{document}

\title{Anatomy of relativistic pion loop corrections to the	\\
	electromagnetic nucleon coupling}

\author{Chueng-Ryong Ji$^1$, W.~Melnitchouk$^2$, A.~W.~Thomas$^3$}
\affiliation{
    $^1$Department of Physics, North Carolina State University,
        Raleigh, North Carolina 27692, USA      \\
    $^2$\mbox{Jefferson Lab, 12000 Jefferson Avenue,
	Newport News, Virginia 23606, USA}	\\
    $^3$CSSM and CoEPP, School of Chemistry and Physics,
        University of Adelaide, Adelaide SA 5005, Australia}

\date{\today}

\begin{abstract}
We present a relativistic formulation of pion loop corrections to the
coupling of photons with nucleons on the light-front.  Vertex and wave
function renormalization constants are computed to lowest order in the
pion field, including their nonanalytic behavior in the chiral limit,
and studied numerically as a function of the ultraviolet cutoff.
Particular care is taken to explicitly verify gauge invariance and
Ward-Takahashi identity constraints to all orders in the $m_\pi$
expansion.  The results are used to compute the chiral corrections
to matrix elements of local operators, related to moments of
deep-inelastic structure functions.  Finally, comparison of results
for pseudovector and pseudoscalar coupling allows the resolution of
a long-standing puzzle in the computation of pion cloud corrections
to structure function moments.
\end{abstract}

\maketitle

\section{Introduction}
\label{sec:intro}

The importance of chiral symmetry in hadron physics has been understood
for more than 50 years.  As an explicit Lagrangian representation of the
approximate chiral symmetry (PCAC) that had been recognized in low energy
pion--nucleon interactions, Gell-Mann and Levy \cite{Gel60} constructed
the extremely successful linear sigma model.  In that model the pion
couples to the nucleon through pseudoscalar coupling, while an additional
scalar ($\sigma$) field is also coupled linearly.  At a more formal
level, Gell-Mann \cite{Gel62} proposed SU(2)$\times$SU(2) as an exact
algebra for the charges associated with the Hamiltonian governing the
strong interaction, even though chiral symmetry was not an exact
symmetry of that Hamiltonian.

On the basis of current algebra one can show very generally that
the amplitude for pion scattering or production must vanish as the
four-momentum of the pion vanishes \cite{Nam62, Adl65}.  Within the
linear sigma model, this important result for low energy pion-nucleon
scattering, for example, is only possible through a subtle cancellation
of two large contributions, the first involving pion emission and
absorption through the pseudoscalar (PS) coupling, and the second
involving $\sigma$ exchange in the $t$-channel.  Keeping track of the
necessary cancellations between such large terms in the linear sigma
model is tedious and for that reason modern formulations of chiral
effective field theory tend to prefer a Lagrangian formulation based
on a nonlinear realization of chiral symmetry
\cite{Wei67, Pag75, Wei79, Gas84}.
In such a formulation the natural $\pi NN$ vertex involves pseudovector
(PV) coupling and the vanishing of pion--nucleon scattering amplitudes
as the pion four-momentum vanishes emerges trivially.  In this work,
motivated by the phenomenological simplicity of enforcing soft-pion
theorems, we focus on the case of PV coupling.  However, since the
linear realization of chiral symmetry is still used in the literature,
for completeness we also compare our results with those for PS coupling.

More recently, chiral symmetry and the pion cloud of the nucleon have
been shown to play a central role in understanding various flavor and
spin asymmetries in quark distribution functions measured at high
energies.
Most prominent of these has been the SU(2) flavor asymmetry in the
proton sea, with the large excess of $\bar d$ quarks over $\bar u$
being predicted in Ref.~\cite{Tho83} and found in deep-inelastic
scattering \cite{NMC} and Drell-Yan experiments \cite{NA51, E866}. 
While a nonperturbative pionic component of the nucleon wave function
provides a natural explanation for the sign of the observed asymmetry,
calculations of the magnitude of the $\bar d - \bar u$ difference have
typically been made in models without a direct connection to QCD.
Furthermore, while the most convenient framework for describing
high-energy reactions is the light-front, the realization of
chiral symmetry on the light-front is yet to be fully understood
(for a recent discussion see, {\it e.g.}, Ref.~\cite{Bea13}).

In Ref.~\cite{JMT09} we examined the framework dependence of pion
loop effects for the simple case of the nucleon self-energy.
We showed that results for the model-independent, nonanalytical
behavior associated with the long-range part of the pion cloud
\cite{LiP71} are in fact independent of whether the calculation
is performed using light-front, instant form (in the rest frame or
infinite momentum frame), or covariant perturbation theory.
On the other hand, important differences were observed for the
nonanalytic structure of the self-energy when comparing the PV
and PS couplings.

Applying the methodologies developed in Refs.~\cite{JMT09, Bur12}, we
consider here the more physically relevant case of the electromagnetic
coupling of the nucleon dressed by pion loops.  This represents the
necessary next step towards the computation of the chiral corrections
to quark distribution functions of the nucleon, whose moments are given
by matrix elements of twist-2 operators.
The twist-2 matrix elements were studied previously by Chen and Ji
\cite{XDJ01} and Arndt and Savage \cite{Sav02}, who computed the most
important pion loop contributions to the leading nonanalytic behavior
within heavy baryon chiral perturbation theory.
In the present analysis we compute the pion loop corrections to the
vertex renormalization factors using a fully relativistic framework,
which includes higher order corrections in the pion mass $m_\pi$.
Furthermore, we demonstrate explicitly that gauge invariance and the
Ward-Takahashi identities hold to all orders $m_\pi$, provided the full
set of one-loop diagrams is considered, including rainbow, tadpole and
Kroll-Ruderman contact terms.
We verify this for both PV and PS theories.
Using the results for the vertex corrections, we then derive the pion
loop corrections to the matrix elements of the twist-2 operators for
both the proton and neutron, and verify the nonanalytic behavior of
the isoscalar and isovector contributions.
Note that for the lowest moment of the nonsinglet distribution the
chiral corrections are essentially those that appear for the nucleon
electromagnetic form factors at zero four-momentum transfer squared,
$q^2=0$.  These have been computed in a relativistic formalism to
one loop order in Refs.~\cite{Kubis01, Fuchs04}.  In contrast to
form factors, for high-energy observables such as quark distribution
functions the natural framework is the light-front, to which we
specialize in this work.  While we also focus on the lowest moment
of the quark distributions, for the reconstruction of the distributions
themselves \cite{DetEPJ01}, higher moments of the distributions will
of course be necessary (Sec.~\ref{sec:me}).

In Sec.~\ref{sec:piN} we review the basics of the pion--nucleon
interaction in terms of the chiral Lagrangian evaluated to lowest
order in derivatives of the pion field.
The electromagnetic nucleon vertex corrections arising from pion
loops are computed in Sec.~\ref{sec:Z1}, and their nonanalytic
properties studied as a function of the pion mass.
To illustrate the role of the various contributions to the vertex
renormalization explicitly, we compute the renormalization factors
numerically as a function of the ultraviolet cutoff.
The results for the vertex corrections and wave function renormalization
are subsequently used in Sec.~\ref{sec:me} to compute the chiral
corrections to nucleon matrix elements of twist-2 operators.
Comparison of the results for PV and PS coupling also allows us
to identify the origin of the discrepancy between the nonanalytic
behaviors of the twist-2 moments computed in heavy baryon chiral
perturbation theory and at the parton level in terms of the Sullivan
process \cite{Sul72}.
Finally, in Sec.~\ref{sec:conc} we summarize our findings
and outline future extensions of the present work.
In Appendix~\ref{app:Feyn} we collect formulas for the complete set
of Feynman rules needed to compute the vertex renormalization and
wave function corrections.  The demonstration that the results respect
gauge invariance and the Ward-Takahashi identity is presented in
Appendices~\ref{app:GI} and \ref{app:WTI}, respectively, and some
useful results for the nonanalytic behavior of integrals are listed
in Appendix~\ref{app:lna}.
Although some of the formal results which we summarize here can be
found elsewhere, our aim will be to provide a pedagogical discussion
of the derivations in order to clarify some conflicting claims in the
literature about the computation of the analytic and nonanalytic
contributions to the pion loop integrals.

\section{Pion-Nucleon Interaction}
\label{sec:piN}

To lowest order in derivatives of the pion field
	$\bm{\pi} = (\pi_+, \pi_-, \pi_0)$,
where $\pi_\pm = (\pi_1 \mp i \pi_2)/\sqrt{2} = \pi_\mp^*$, the
$\pi NN$ Lagrangian is given by \cite{Wei67, Kra90, Jen91, BKM95}
\begin{eqnarray}
{\cal L}_{\pi N}
&=& {g_A \over 2 f_\pi}\,
    \bar\psi_N \gamma^\mu \gamma_5\,
	\bm{\tau} \cdot \partial_\mu \bm{\pi}\, \psi_N\
 -\ {1 \over (2 f_\pi)^2}\,
    \bar\psi_N \gamma^\mu\, \bm{\tau} \cdot
	(\bm{\pi} \times \partial_\mu \bm{\pi})\, \psi_N,
\label{eq:LpiN}
\end{eqnarray}
where $\psi_N$ is the nucleon field, $\vec\tau$ is the Pauli matrix
operator in nucleon isospin space, $f_\pi = 93$~MeV is the pion decay
constant, and $g_A = 1.267$ is the nucleon axial vector charge.
Our convention follows that in Ref.~\cite{Jen91}, but differs by an
overall minus sign for the $g_A$ term in Eq.~(\ref{eq:LpiN}) from
that in Ref.~\cite{BKM95}.  For quantities where the $\bm{\pi}$ field
enters quadratically, such as the pion loop corrections discussed
here, the overall sign on the $g_A$ term is immaterial.
The $g_A$-dependent term in the Lagrangian (\ref{eq:LpiN}) gives rise
to the ``rainbow'' diagram in which a pion is emitted and absorbed by
the nucleon at different space-time points, while the second is the
Weinberg-Tomozawa coupling \cite{Wei67, Tom66}, which has two pion
fields coupling to the nucleon at the same point.  The latter gives the
leading contribution to S-wave pion-nucleon scattering \cite{Wei66},
and generates the pion tadpole or bubble diagrams.

The above definitions mean that the field $\pi_-$ corresponds to
an incoming negatively charged pion, with $\pi_+^*$ to an outgoing
positively charged pion.
In writing Eq.~(\ref{eq:LpiN}) we have also made use of the
Goldberger-Treiman relation between $g_A$, $f_\pi$ and the
$\pi NN$ coupling constant $g_{\pi NN}$,
\begin{eqnarray}    
{g_A \over f_\pi} &=& {g_{\pi NN} \over M} ,
\label{eq:GT}
\end{eqnarray}
where $g_{\pi NN} \approx 13.4$ and $M$ is the nucleon mass.

The interaction of pions and nucleons with the electromagnetic
field is introduced by minimal substitution,
$\partial_\mu \to \partial_\mu + i e A_\mu$, where the charge
$e = -1$ for a photon coupling to an electron.
This gives rise to a $\gamma\pi N$ interaction Lagrangian
of the form
\begin{eqnarray}
{\cal L}_{\gamma\pi N}
&=& - \bar\psi_N \gamma_\mu\, \hat{Q}_N \psi_N\, A_\mu\
 +\ i\, (\partial^\mu \bm{\pi}) \cdot (\hat{Q}_\pi \bm{\pi})\, A_\mu
							\nonumber\\
&+& {i g_A \over 2 f_\pi}\,
    \bar\psi_N \gamma^\mu \gamma_5\,
	\bm{\tau} \cdot \hat{Q}_\pi \bm{\pi}\, \psi_N\, A_\mu\
 -\ {i \over (2 f_\pi)^2}\,
    \bar\psi_N \gamma^\mu\, \bm{\tau} \cdot
	(\bm{\pi} \times \hat{Q}_\pi \bm{\pi})\, \psi_N\, A_\mu
\label{eq:LpiNg}
\end{eqnarray}
where $\hat{Q}_N = |e| (I+\tau_3)/2$ is the nucleon charge operator,
defined in terms of the total isospin $I$ of the nucleon and its
third component such that
$\hat{Q}_p \psi_p = |e| \psi_p$, and
$\hat{Q}_n \psi_n = 0$,
and similarly for the pion charge operator one has
$\hat{Q}_\pi \bm{\pi} = |e| (\pi_+, -\pi_-, 0)$.
The first two terms in Eq.~(\ref{eq:LpiNg}) correspond to the photon
coupling to the bare nucleon and pion, respectively, the third term
is the Kroll-Ruderman coupling \cite{Kro54, Dre92} required by gauge
invariance, while the fourth term gives rise to a photon coupling to
a pion--nucleon tadpole vertex.
The Lagrangians (\ref{eq:LpiN}) and (\ref{eq:LpiNg}) can be used
to derive a set of Feynman rules for computing lowest order
amplitudes, which are summarized in Appendix~\ref{app:Feyn}.

The transformation of the PS coupling Lagrangian, such as in the
linear $\sigma$ model \cite{Gel60, Sch57}, to the PV coupling
Lagrangian in Eq.~(\ref{eq:LpiN}) can be understood as a canonical
transformation of the field variables, analogous to the coordinate
transformation from the Cartesian coordinates $x$ and $y$ to the
plane polar coordinates $r = \sqrt{x^2+y^2}$ and
$\theta = \tan^{-1}\left(y/x\right)$ \cite{Wei67, Wal95}.
In analogy with the coordinate transformation, the PV coupling
Lagrangian can be derived from the PS Lagrangian with the form
of $x+iy$ (representing, for example,
$M - g_{\pi N N}\, \sigma
 - i g_{\pi N N}\, \gamma_5\, \bm{\tau} \cdot \bm{\pi}$)
by taking the equivalence between the ($x, y$) and ($r, \theta$)
coordinates, such that $x+iy = r e^{i\theta}$ with an appropriate
field redefinition for the nucleon, pion and $\sigma$ fields.
Since the redefined scalar field turns out to be completely decoupled
after the canonical transformation and becomes irrelevant to the
chiral symmetry, one may completely remove it in the PV coupling
Lagrangian \cite{Wei67, Kra90, Jen91, BKM95, Tho81}, as shown in
Eqs.~(\ref{eq:LpiN}) and (\ref{eq:LpiNg}).
In the original PS coupling Lagrangian it is crucial
to keep the $\sigma$ field to maintain the chiral symmetry.
Thus, the pseudoscalar Lagrangian is not invariant under
chiral transformations without the presence of a scalar field,
as in the linear sigma model~\cite{TW01}.

However, historically the pseudoscalar pion--nucleon interaction
without a scalar field has often been discussed in the literature,
with the lowest order Lagrangian density given by
\begin{eqnarray}
{\cal L}^{\rm PS}_{\pi N}
&=& -g_{\pi NN}\,
    \bar\psi_N\, i \gamma_5\, \bm{\tau} \cdot \bm{\pi}\, \psi_N\
        +\ \cdots
\label{eq:Lps}
\end{eqnarray}
As discussed by Lensky and Pascalutsa \cite{Lensky09}, this can be
formally obtained by redefining the nucleon field
$\psi_N \to \xi \psi_N$, where
  $\xi = \exp[ (i g_A {\bm{\tau} \cdot \bm{\pi}} / 2 f_\pi) \gamma_5]$,
which leads to the PS Lagrangian in Eq.~(\ref{eq:Lps}), together with
the Weinberg-Tomozawa contribution in (\ref{eq:LpiN}) replaced by an
isoscalar term as in the $\sigma$ model and a modified isovector term.

To contrast a number of important features pertinent to the PV and PS
calculations, we discuss here the consequences of neglecting the scalar
field contribution for the PS theory.  For on-shell nucleons obeying
the free Dirac equation, the PS and PV Lagrangians (\ref{eq:LpiN})
give identical results for matrix elements, provided the couplings
are related by Eq.~(\ref{eq:GT}).  For off-shell nucleons, however,
the PS and PV interactions lead to different results because of the
strong coupling to negative energy states in the former.  This can be
illustrated by splitting the nucleon off-shell propagator into an
on-shell part and an off-shell part, according to the identity
\cite{Cha73}
\begin{equation}
\frac{1}{p\!\!\!\slash - M} 
= \frac{\sum_s u(p,s){\bar u}(p,s)}{p^2-M^2}
+ \frac{\gamma^+}{2p^+}, 
\end{equation}
where $\gamma^+ = \gamma^0 + \gamma^3$ and $p^+ = p_0 + p_z$.
One observes that while the on-shell component gives equivalent
results for PV and PS interactions, the contribution from the off-shell
part $\gamma^+ / 2p^+$ differs for the PV and PS couplings \cite{JMT13}.
These differences were studied in detail for the case of the nucleon
self-energy $\Sigma$ in Ref.~\cite{JMT09}, where the leading nonanalytic
behavior of $\Sigma$ was found to be of order $m_\pi^2 \log m_\pi^2$ for
the PS case, in contrast to the $m_\pi^3$ behavior of the PV theory.
In the nucleon self-energy, the difference in the off-shell part
$\gamma^+ / 2p^+$ indeed appears as a pion tadpole contribution in
the PS theory.  Since the pion tadpole involves the two-pion coupling,
which corresponds to a scalar coupling, it is evident that the
equivalence between the PV and PS coupling theories cannot be
attained without a scalar field to restore the chiral symmetry.
Important differences arise also for the vertex corrections,
as we shall discuss in the following.

\section{Vertex Corrections}
\label{sec:Z1}

\begin{figure}[tb]
\includegraphics[width=12cm]{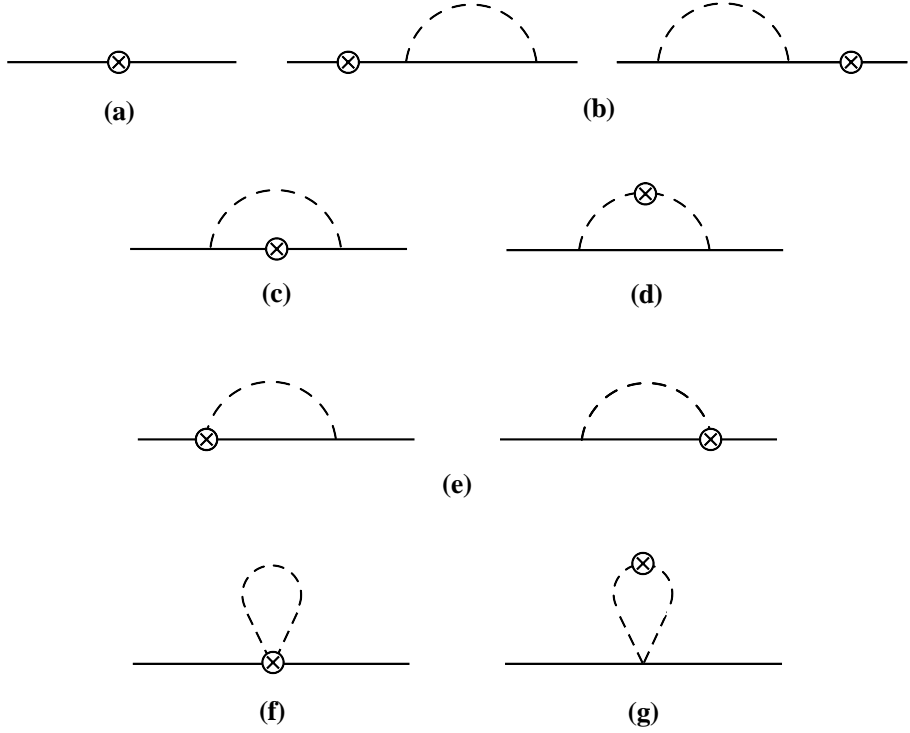}
\caption{Pion loop corrections to the photon--nucleon coupling
	in the PV pion-nucleon theory:
	(a) photon coupling to the bare nucleon,
	(b) wave function renormalization,
	(c) rainbow diagram with coupling to the nucleon,
	(d) rainbow diagram with coupling to the pion,
	(e) Kroll-Ruderman diagrams,
	(f) pion tadpole diagram with coupling to the pion--nucleon vertex,
	(g) pion bubble diagram with coupling to the pion.}
\label{fig:loops}
\end{figure}

Beyond tree level, the interactions described by Eqs.~(\ref{eq:LpiN})
and (\ref{eq:LpiNg}) give rise to loop corrections which renormalize
the electromagnetic photon--nucleon vertex.  These corrections are
illustrated in Fig.~\ref{fig:loops}.  In this section we will derive
the corrections arising from each of these diagrams explicitly and
study their dependence on the high-momentum cutoff mass, as well as
their nonanalytic properties as a function of the pion mass.
For illustration, we estimate the contribution from each diagram
numerically by introducing ultraviolet regularization cutoff
parameters in this work.  More quantitative numerical estimates
using Lorentz invariant regularization methods, such as dimensional
regularization, will be presented in future work \cite{CRJ13}.

The vertex renormalization constant $Z_1$ is defined as
\begin{eqnarray}
(Z_1^{-1} - 1)\, \bar u(p)\, (-i |e| \gamma^\mu)\, u(p)
&=& (-i |e|)\, \bar u(p)\, \Lambda^\mu\, u(p),
\label{eq:Z1def}
\end{eqnarray}
where the operator $\Lambda^\mu$ is given by the sum of vertex
correction diagrams in Fig.~\ref{fig:loops}(c)--(g), and for
convenience is defined with the charge factor $(-i |e|)$ taken out.
We use the convention in which the nucleon spinors are normalized
according to $\bar u(p) u(p) = 1$.
For small values of $Z_1$, one also has, to lowest order in the
$\pi N$ coupling, $Z_1^{-1} - 1 \approx 1 - Z_1$.
To evaluate the vertex renormalization constants from the various
diagrams in Fig.~\ref{fig:loops}, we take the $\mu=+$ components,
so that
\begin{eqnarray}
1 - Z_1
&=& {M \over p^+}\, \bar u(p)\, \Lambda^+\, u(p),
\label{eq:Lam+}
\end{eqnarray}
where the $\pm$ components of the momentum four-vector are given
by $p^\pm = p_0 \pm p_z$.

\subsection{Photon-nucleon coupling}
\label{ssec:Z1N}

For the coupling of the photon to a proton dressed by a neutral
$\pi^0$ loop ($p \to p + \pi^0$), Fig.~\ref{fig:loops}(c),
the vertex correction is computed from the operator
\begin{eqnarray}
\Lambda^\mu_p
&=& \left( {g_A \over 2 f_\pi} \right)^2
\int\!\!{d^4k \over (2\pi)^4}\,
	(k\!\!\!\slash \gamma_5)\,
	{i (p\!\!\!\slash - k\!\!\!\slash + M) \over D_N}\,
	\gamma^\mu\,
	{i (p\!\!\!\slash - k\!\!\!\slash + M) \over D_N}\,
	(\gamma_5 k\!\!\!\slash)\,
	{i \over D_\pi},
\label{eq:LamN}
\end{eqnarray}
where we use the shorthand notation for the pion and nucleon
propagators,
\begin{subequations}
\begin{eqnarray}
D_\pi &\equiv& k^2 - m_\pi^2 + i\eps,		\\
D_N   &\equiv& (p-k)^2 - M^2 + i\eps,
\end{eqnarray}
\end{subequations}
respectively.  For a neutron target, the correction from the coupling
to an intermediate state proton dressed by a negatively charged pion
($n \to p + \pi^-$) is given by $\Lambda^\mu_n = 2 \Lambda^\mu_p$.
Taking the $\mu = +$ component and using Eq.~(\ref{eq:Lam+}),
the contribution to the vertex renormalization factor
$(1-Z_1^p) = {1\over 2} (1-Z_1^n) \equiv (1-Z_1^N)$ is then given by
\begin{eqnarray}
1 - Z_1^N
&=& i \left( {g_A \over 2 f_\pi} \right)^2
\int\!\!{d^4k \over (2\pi)^4}
  \left[ k^4 + 4 (p\cdot k)^2 - 4 M^2 k^2 (1-y) - 4 p\cdot k\, k^2
  \right]
  {1 \over D_\pi D_N^2 }				\nonumber\\
&=& -i \left( {g_A \over 2 f_\pi} \right)^2
\int\!\!{d^4k \over (2\pi)^4}
\left[ {4 M^2 (k^2 - 2y\, p\cdot k) \over D_\pi D_N^2}\
    -\ {4 M^2 y \over D_\pi D_N}\
    -\ {1 \over D_\pi}
\right],
\label{eq:Z1N}
\end{eqnarray}
where $y = k^+/p^+$ is the fraction of the nucleon's $+$ component
of momentum carried by the pion.
In deriving Eq.~(\ref{eq:Z1N}) we have used the Dirac equation,
$p\!\!\!\slash u(p) = M u(p)$.

Since there are two more powers of momentum $k$ in the numerator
of $Z_1^N$ than in the denominator, the integral (\ref{eq:Z1N})
is formally divergent.  One can perform the loop integration
and regularize the divergence in several ways.
In Ref.~\cite{JMT09} we considered the nucleon self-energy arising
from pion dressing, and computed the loop integrals using equal-time
perturbation theory in the rest frame and in the infinite momentum
frame, using light-front coordinates (all with appropriate
high-momentum cutoffs), and covariantly with dimensional
regularization.  Each method was shown to give identical results
for the model-independent, nonanalytic part of the integrals,
with the (model-dependent) analytic contributions dependent
upon the regularization prescription.

In the case of the vertex renormalization, it is convenient to use
the $+$ prescription to evaluate $Z_1$, and it will be particularly
instructive to examine the integrands as a function of $k^+$ (or $y$)
and $k_\perp$.  Performing the $k^-$ integration using the Cauchy
integral theorem by closing the contour in the lower half-plane, one
can write the nucleon contribution to the vertex renormalization as
\begin{eqnarray}
1 - Z_1^N
&=& {g_A^2 M^2 \over (4\pi f_\pi)^2}
\int\!dy\, dk_\perp^2
\left\{ { y (k_\perp^2 + y^2 M^2) \over
	  \left[ k_\perp^2 + y^2 M^2 + (1-y) m_\pi^2 \right]^2 }\
     -\ { y \over k_\perp^2 + y^2 M^2 + (1-y) m_\pi^2 }\
\right.						\nonumber\\
& & \hspace*{3.5cm}
\left.
     -\ { 1 \over 4M^2 }
	\log\left( {k_\perp^2+m_\pi^2 \over \mu^2} \right)
	\delta(y)
\right\},
\label{eq:Z1Ny}
\end{eqnarray}
where the mass $\mu$ is an ultraviolet cutoff on the $k^-$ integration.
In Eq.~(\ref{eq:Z1Ny}) the first, second and third terms correspond
to the terms in the integrand of Eq.~(\ref{eq:Z1N}) proportional
to $1/D_\pi D_N^2$, $1/D_\pi D_N$ and $1/D_\pi$, respectively.
The first term, which is logarithmically divergent in $k_\perp$,
gives a contribution that is equivalent to that obtained from a
PS pion--nucleon coupling (\ref{eq:Lps}) \cite{DLY70, TMS00},
\begin{eqnarray}
1 - \widetilde{Z}_1^N
&=& -i g_{\pi NN}^2
\int\!\!{d^4k \over (2\pi)^4}\,
	{k^2 - 2y\, p\cdot k \over D_\pi D_N^2}	\nonumber\\
&=& {g_{\pi NN}^2 \over 16\pi^2}
\int\!dy\, dk_\perp^2\,
{ y\, (k_\perp^2 + y^2 M^2) \over
  \left[ k_\perp^2 + y^2 M^2 + (1-y) m_\pi^2 \right]^2 },
\label{eq:Z1Nps}
\end{eqnarray}
where the couplings are related as in Eq.~(\ref{eq:GT}), and we use
the tilde (`` $^{\sim}$ '') notation for quantities computed from
the PS interaction.  This result agrees exactly with the result
from the infinite momentum frame calculation of Drell, Levy and
Yan \cite{DLY70} within the PS $\pi N$ theory.

\begin{figure}[t]
\includegraphics[width=10cm]{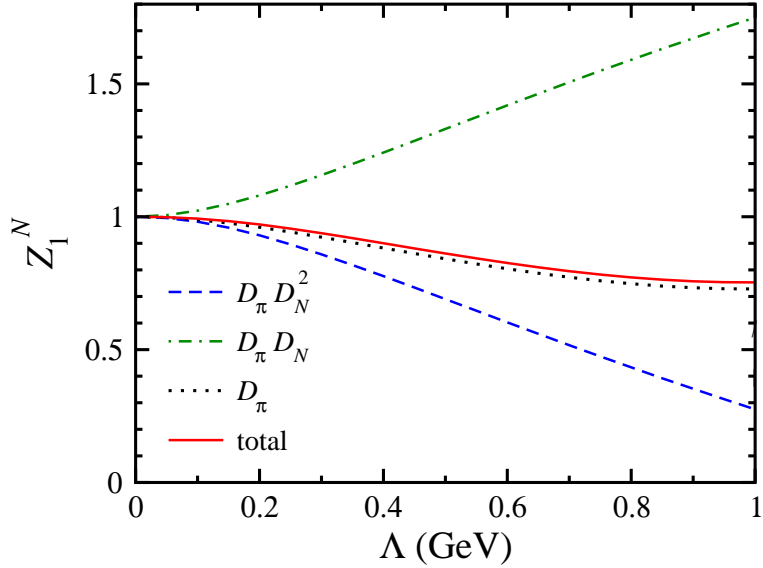}
\caption{Contributions to the vertex renormalization $Z_1^N$
	from terms in Eq.~(\ref{eq:Z1N}) proportional to
	$1/D_\pi D_N^2$ (dashed),
	$1/D_\pi D_N$ (dot-dashed),
	$1/D_\pi$ (dotted),
	and the sum (solid), as a function of the $k_\perp$
	momentum cutoff $\Lambda$.}
\label{fig:Z1N}
\end{figure}

The second term in Eq.~(\ref{eq:Z1Ny}) is a new contribution,
associated with the momentum dependence of the PV pion--nucleon
vertex, and enters with the opposite sign to the PS-like component.
The third term is nonzero only at $y=0$, and arises from the
$1/D_\pi$ term in Eq.~(\ref{eq:Z1N}).
For $k_\perp \lesssim \mu$ this serves to enhance the contribution
from the $1/D_\pi D_N^2$ PS-like term.

Numerically, the contributions to $Z_1^N$ from each of the three
terms are illustrated in Fig.~\ref{fig:Z1N} as a function of the
cutoff mass $\Lambda$ used to render the $k_\perp$ integration
finite, taking the $k^-$ integration cutoff $\mu = 1$~GeV.
The results show large cancellation between the $1/D_\pi D_N^2$
term (which gives a negative contribution to $1-Z_1^N$) and the
$1/D_\pi D_N$ term (which gives a positive contribution), with
the total closely following the residual $1/D_\pi$ contribution,
which is smaller in magnitude than the other two pieces.
This clearly illustrates that a calculation of the vertex
renormalization using the PS $\pi N$ interaction, apart from
not having the correct chiral symmetry properties, yields very
different results phenomenologically compared with the PV theory.

\subsection{Photon-pion coupling}
\label{ssec:Z1pi}

For the photon coupling to a positively charged pion emitted from the
proton, Fig.~\ref{fig:loops}(d), the corresponding operator can be
written
\begin{eqnarray}
\Lambda^\mu_{\pi^+}
&=& 2 \left( {g_A \over 2 f_\pi} \right)^2
\int\!\!{d^4k \over (2\pi)^4}
	(k\!\!\!\slash \gamma_5)
	{i (p\!\!\!\slash - k\!\!\!\slash + M) \over D_N}\,
	(\gamma_5 k\!\!\!\slash)
	{i \over D_\pi}
	{i \over D_\pi}\,
	(2k^\mu),
\label{eq:Lmupi}
\end{eqnarray}
where the overall isospin factor 2 accounts for the $p \to n \pi^+$
transition.  For a neutron target, the operator for the coupling
to the negatively charged pion in the transition $n \to p \pi^-$
would be $\Lambda^\mu_{\pi^-} = -\Lambda^\mu_{\pi^+}$.
Taking the $\mu = +$ component in Eq.~(\ref{eq:Lmupi}), the
resulting vertex renormalization factor for the pion coupling
$(1-Z_1^{\pi^+}) = -(1-Z_1^{\pi^-}) \equiv 2(1-Z_1^\pi)$
can be written as
\begin{eqnarray}
1 - Z_1^\pi
&=& i \left( {g_A \over 2 f_\pi} \right)^2
\int\!\!{d^4k \over (2\pi)^4}
  \left[ 2 (p\cdot k)^2 - k^2 p\cdot k - 2 M^2 k^2 
  \right]
  {2y \over D_\pi^2 D_N}				\nonumber\\
&=& -i \left( {g_A \over 2 f_\pi} \right)^2
\int\!\!{d^4k \over (2\pi)^4}
\left[ {8y M^2\, p\cdot k \over D_\pi^2 D_N}\
    +\ {2y\, p \cdot k    \over D_\pi^2}
    +\ {4y\, M^2          \over D_\pi^2}
\right],
\label{eq:Z1pi}
\end{eqnarray} 
where $Z_1^\pi$ here is defined with the isospin factor removed.
Because $y$ is odd in $k^+$, while $D_\pi^2$ is even, the third
term in Eq.~(\ref{eq:Z1pi}) proportional to $y/D_\pi^2$ will vanish
after integration over $k^+$.
For the second term, proportional to $2y p \cdot k / D_\pi^2$,
we can use the identity
\begin{eqnarray}
\int\!\!{d^4k}\, {2y\, p\cdot k \over D_\pi^2}
&=& \int\!\!{d^4k} {1 \over D_\pi}.
\label{eq:identity}
\end{eqnarray}
In fact, since
\begin{eqnarray}
{\partial \over \partial k^+} {1 \over D_\pi^n}
&=& -{n k^- \over D_\pi^{n+1}},
\end{eqnarray}
one has, for any integer $n$,
\begin{eqnarray}
\int\!\!{d^4k}\, {2y\, p\cdot k \over D_\pi^{n+1}}
&=& {1 \over n} \int\!\!{d^4k} {1 \over D_\pi^n}.
\end{eqnarray}
Using the relation (\ref{eq:identity}), and performing the $k^-$
integration by closing the contour in the upper half-plane,
the photon--pion coupling contribution to $Z_1$ can be written as
\begin{eqnarray}
1 - Z_1^\pi
&=& {g_A^2 M^2 \over (4\pi f_\pi)^2}
\int\!dy\, dk_\perp^2
\left\{ { y (k_\perp^2 + y^2 M^2) \over
          \left[ k_\perp^2 + y^2 M^2 + (1-y) m_\pi^2 \right]^2 }
     +\ { 1 \over 4M^2 }
        \log\left( {k_\perp^2+m_\pi^2 \over \mu^2} \right) \delta(y)
\right\}.					\nonumber\\
& &
\label{eq:Z1piy}
\end{eqnarray}
Note that the first term in Eq.~(\ref{eq:Z1piy}) is identical to
the first term in Eq.~(\ref{eq:Z1Ny}) for the $Z_1^p$ contribution
to the vertex renormalization.  It is also the result one obtains
from the ``Sullivan process'' \cite{Sul72, Tho83, Hen90, Zol91, Mel93,
Hol96, Spe97, Kum98} for the contribution of the pion cloud to the
deep-inelastic structure function of the nucleon, where the current
couples to the pion cloud, leaving an on-shell nucleon in the final
state.  These analyses all utilized the PS pion--nucleon interaction,
in which the vertex renormalization factor is given by
\begin{eqnarray}
1 - \widetilde{Z}_1^\pi
&=& -i g_{\pi NN}^2\,
\int\!\!{d^4k \over (2\pi)^4}\,
	{2y\, p\cdot k \over D_\pi^2 D_N}	\nonumber\\
&=& {g_{\pi NN}^2 \over 16\pi^2}
\int\!dy\, dk_\perp^2\,
{ y\, (k_\perp^2 + y^2 M^2) \over
        \left[ k_\perp^2 + y^2 M^2 + (1-y) m_\pi^2 \right]^2 }.
\label{eq:Z1piyPS}
\end{eqnarray}
This result also coincides with the vertex renormalization computed
in the infinite momentum frame in Ref.~\cite{DLY70} in terms of
nucleon and pion ``partons'' in the PS theory.
Comparison of Eqs.~(\ref{eq:Z1piyPS}) and (\ref{eq:Z1Nps}) also
demonstrates that the PS model respects the charge conservation
condition,
\begin{eqnarray}
1 - \widetilde{Z}_1^\pi &=& 1 - \widetilde{Z}_1^N,
\label{eq:PSgi}
\end{eqnarray}
which follows directly from the Ward-Takahashi identity
(see Appendix~\ref{app:WTI}).

\begin{figure}[tb]
\includegraphics[width=10cm]{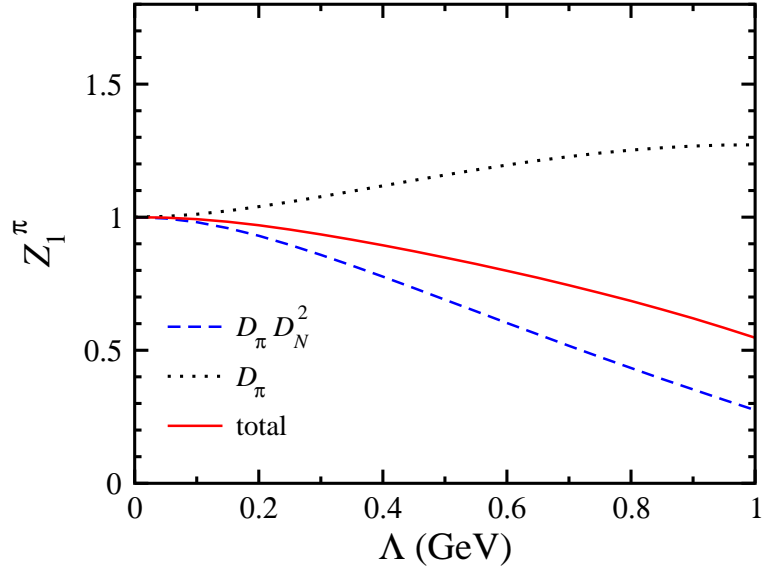}
\caption{Contributions to the vertex renormalization $Z_1^\pi$
	from terms in Eq.~(\ref{eq:Z1piy}) proportional to
	$1/D_\pi D_N$ (dot-dashed),
	$1/D_\pi$ (dotted),
	and the sum (solid), as a function of the $k_\perp$
	momentum cutoff $\Lambda$.}
\label{fig:Z1pi}
\end{figure}

In contrast, the full PV result for $Z_1^\pi$ in Eq.~(\ref{eq:Z1piy})
contains in addition a singular, $\delta$-function term in $y$, just
as in Eq.~(\ref{eq:Z1Ny}) for $Z_1^p$ but with the opposite sign.
As illustrated in Fig.~\ref{fig:Z1pi}, this term cancels some of the
contribution from the PS, $1/D_\pi D_N^2$ term, leaving an overall
positive contribution to $1-Z_1^\pi$.
Clearly the PV $Z_1^\pi$ and $Z_1^p$ results are different for any
value of the $k_\perp$ cutoff $\Lambda$, and in order to demonstrate
their equivalence requires consideration of additional terms arising
from the derivative PV coupling.

\subsection{Kroll-Ruderman terms}
\label{ssec:Z1kr}

The momentum dependence of the PV $\pi N$ interaction gives rise to an
additional Kroll-Ruderman (KR) term \cite{Tom66} which describes the
photon coupling to the $\pi NN$ vertex, Fig.~\ref{fig:loops}(e).
For the case of the $p \to n \pi^+$ vertex, these are computed
from the operator
\begin{eqnarray}
\Lambda^\mu_{{\rm KR}, \pi^+}
&=& 2 i \left( {g_A \over 2 f_\pi} \right)^2
\int\!\!{d^4k \over (2\pi)^4}
 \left( k\!\!\!\slash \gamma_5
        {i (p\!\!\!\slash - k\!\!\!\slash + M) \over D_N}
        \gamma^\mu \gamma_5\
     +\ \gamma_5 \gamma^\mu  
        {i (p\!\!\!\slash - k\!\!\!\slash + M) \over D_N}
        \gamma_5 k\!\!\!\slash
 \right)
        {i \over D_\pi}.			\nonumber\\
& &
\label{eq:Lmukr}
\end{eqnarray}
For a neutron target, the corresponding operator describing
the photon coupling to the $n \to p \pi^-$ vertex is
$\Lambda^\mu_{{\rm KR}, \pi^-} = - \Lambda^\mu_{{\rm KR}, \pi^+}$.
Note that the emission of a neutron $\pi^0$ does not give rise
to a KR correction term.

The contribution to the vertex renormalization factor is
then given by
$(1 - Z_1^{{\rm KR},\pi^+}) = - (1 - Z_1^{{\rm KR},\pi^-})
                       \equiv 2 (1 - Z_1^{\rm KR})$,
where
\begin{eqnarray}   
1 - Z_1^{\rm KR}
&=& -i \left( {g_A \over 2 f_\pi} \right)^2
\int\!\!{d^4k \over (2\pi)^4}
  \left( 2 k^2 - 4 p \cdot k + 4 M^2 y \right)
  {1 \over D_\pi D_N}				\nonumber\\
&=& -i \left( {g_A \over 2 f_\pi} \right)^2
\int\!\!{d^4k \over (2\pi)^4}
\left( -{4 M^2 y \over D_\pi D_N}\ -\ {2 \over D_\pi}
\right).
\label{eq:Z1kr}
\end{eqnarray}
Performing the $k^-$ integration, the resulting contribution to
$1-Z_1$ is
\begin{eqnarray}
1 - Z_1^{\rm KR}
&=& -{g_A^2 M^2 \over (4\pi f_\pi)^2}
\int\!dy\, dk_\perp^2\,
\left\{    {y \over k_\perp^2 + y^2 M^2 + (1-y) m_\pi^2}
        +\ { 1 \over 2M^2 }
	   \log\left( {k_\perp^2+m_\pi^2 \over \mu^2} \right)
	   \delta(y)
\right\},					\nonumber\\
& &
\label{eq:Z1kry}
\end{eqnarray}
where the first term in the integrand arises from the $1/D_\pi D_N$
term, while the $\delta(y)$ term is associated with the $1/D_\pi$
contribution in Eq.~(\ref{eq:Z1kr}).

\begin{figure}[tb]
\includegraphics[width=10cm]{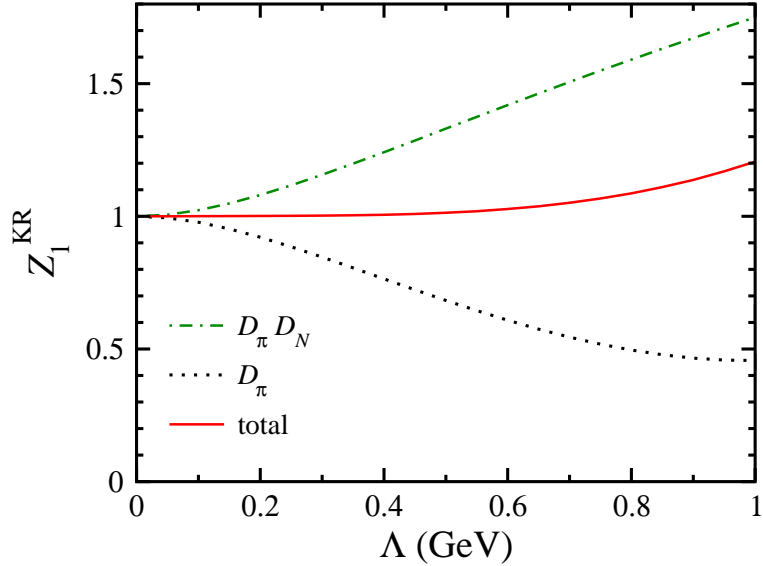}
\caption{Contributions to the vertex renormalization $Z_1^{\rm KR}$
	from terms in Eq.~(\ref{eq:Z1kry}) proportional to
	$1/D_\pi D_N$ (dot-dashed),
	$1/D_\pi$ (dotted),
	and the sum (solid), as a function of the $k_\perp$
	momentum cutoff $\Lambda$.}
\label{fig:Z1kr}
\end{figure}

The contributions from the individual terms to $Z_1^{\rm KR}$ are
shown in Fig.~\ref{fig:Z1kr} as a function of the $k_\perp$
momentum cutoff $\Lambda$, as well as the total KR correction.
Note the large cancellation between the $1/D_\pi$ and $1/D_\pi D_N$
terms for values of the cutoff $\Lambda \lesssim 0.8$~GeV.
Omission of the $\delta$-function $1/D_\pi$ contribution would
thus lead to a significant overestimate of the KR correction.

\begin{figure}[tb]
\includegraphics[width=10cm]{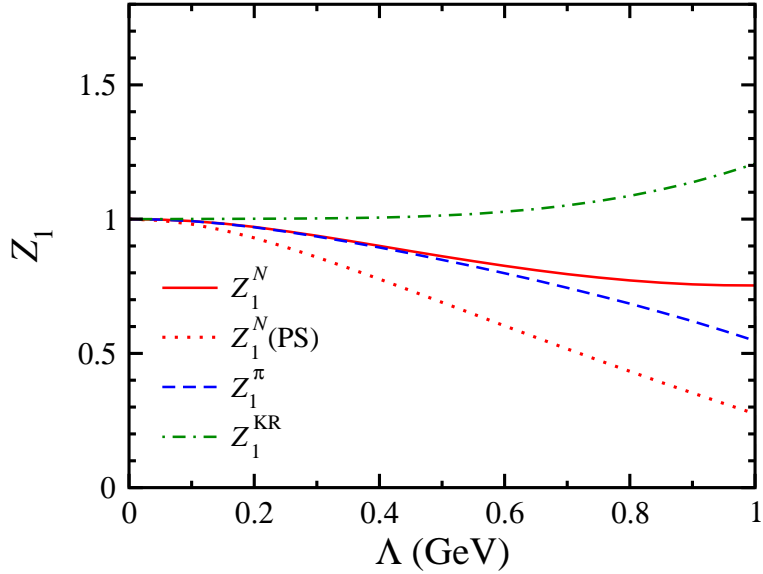}
\caption{Contributions to the vertex renormalization $Z_1$
	from the photon--nucleon coupling $Z_1^N$ (solid),
	photon--pion coupling $Z_1^\pi$ (dashed), and
	Kroll-Ruderman terms $Z_1^{\rm KR}$ (dot-dashed),
	as a function of the $k_\perp$ momentum cutoff $\Lambda$.
	Note that the sum of the pion tadpole and bubble
	contributions to $Z_1$ vanishes.}
\label{fig:Z1}
\end{figure}

Formally, the Kroll-Ruderman terms are needed to ensure gauge
invariance in the PV theory.  Indeed, from Eqs.~(\ref{eq:Z1Ny}),
(\ref{eq:Z1piy}) and (\ref{eq:Z1kry}) one can verify explicitly
that
\begin{eqnarray}
(1-Z_1^N) &=& (1-Z_1^\pi) + (1-Z_1^{\rm KR})
\label{eq:PVgi}
\end{eqnarray}
for the PV case.  In contrast, in the PS theory, where the $\pi N$
vertex is independent of momentum, there is no analogous KR
contribution, and gauge invariance is reflected through the relation
(\ref{eq:PSgi}).
In Fig.~\ref{fig:Z1} we show the total contributions to $Z_1$ from
the nucleon and pion rainbow diagrams, $Z_1^N$ and $Z_1^\pi$, and 
the total KR correction.  The sum of these three terms is of course
zero by Eq.~(\ref{eq:PVgi}).

\subsection{Tadpoles and bubbles}
\label{ssec:Z1tad}

At lowest order the Lagrangian (\ref{eq:LpiNg}) contains,
in addition to the PV coupling of the pion to the nucleon,
quadratic terms arising from the covariant derivative.
For the coupling of the photon to the $\pi\pi pp$ vertex
in the pion tadpole diagram in Fig.~\ref{fig:loops}(f)
the relevant operator is
\begin{eqnarray}
\Lambda^\mu_{p\, {\rm tad}}
&=& -{1 \over 2f_\pi^2}
\int\!\!{d^4k \over (2\pi)^4}\,
	\gamma^\mu\,
	{i \over D_\pi}.
\label{eq:LmuNtad}
\end{eqnarray}
For the coupling to the $\pi\pi nn$ vertex, the corresponding operator
is $\Lambda^\mu_{n\, {\rm tad}} = -\Lambda^\mu_{p\, {\rm tad}}$.
The contribution to the vertex renormalization from the $\pi\pi NN$
tadpoles is then
$(1 - Z_1^{N\, {\rm tad}})
\equiv (1 - Z_1^{p\, {\rm tad}})
   = - (1 - Z_1^{n\, {\rm tad}})$,
where
\begin{eqnarray}
1 - Z_1^{N\, {\rm tad}}
&=& -{i \over 2 f_\pi^2}
\int\!\!{d^4k \over (2\pi)^4} {1 \over D_\pi}.
\label{eq:Z1Ntad}
\end{eqnarray}
After integration over $k^-$, this can be written
\begin{eqnarray}
1 - Z_1^{N\, {\rm tad}}
&=& {1 \over 2 (4\pi f_\pi)^2}
\int\!dy\, dk_\perp^2\,
\log\left( {k_\perp^2+m_\pi^2 \over \mu^2} \right) \delta(y).
\label{eq:Z1Ntady}
\end{eqnarray}

\begin{figure}[tb]
\includegraphics[width=10cm]{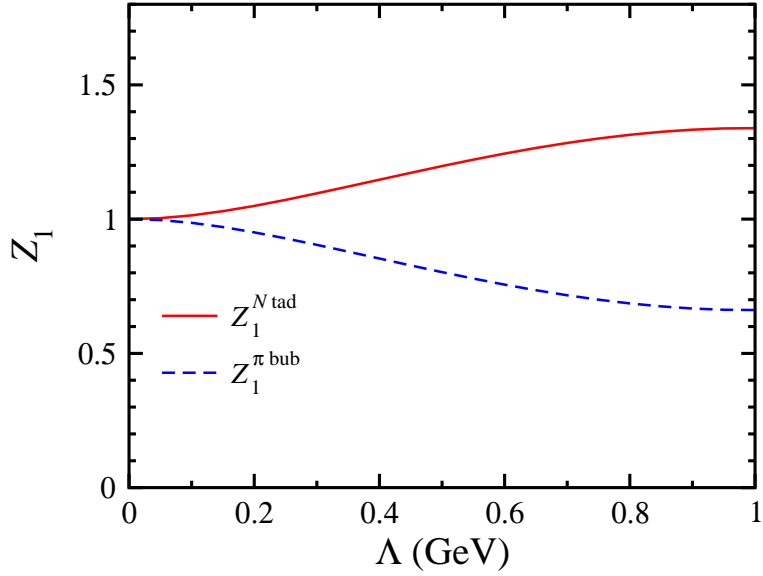}
\caption{Nucleon tadpole (solid) and pion bubble (dashed)
	contributions to the vertex renormalization as a function
	of the $k_\perp$ momentum cutoff $\Lambda$.}
\label{fig:Z1tad}
\end{figure}

For the bubble diagram with the photon coupling directly to a pion,
Fig.~\ref{fig:loops}(g), the relevant operator for a proton target is
\begin{eqnarray}
\Lambda^\mu_{\pi\, {\rm bub} (p)}
&=& {1 \over 2f_\pi^2}
\int\!\!{d^4k \over (2\pi)^4}\,
	(-i k\!\!\!\slash)\,
	2 k^\mu\,
	{i \over D_\pi}
	{i \over D_\pi},
\label{eq:Lmupibub}
\end{eqnarray}
with that for a neutron target given by
$\Lambda^\mu_{\pi\, {\rm bub} (n)}
= -\Lambda^\mu_{\pi\, {\rm bub} (p)}$.
Taking the $\mu = +$ component on both sides of Eq.~(\ref{eq:Lmupibub}),
the contribution to the vertex renormalization is given by
$(1-Z_1^{\pi\, {\rm bub}}) \equiv (1-Z_1^{\pi\, {\rm bub} (p)})
			      = - (1-Z_1^{\pi\, {\rm bub} (n)})$,
where
\begin{eqnarray}
1 - Z_1^{\pi\, {\rm bub}}
&=& {i \over 2 f_\pi^2}
\int\!\!{d^4k \over (2\pi)^4}\, {2 y\, p \cdot k \over D_\pi^2}.
\label{eq:Z1pibub}
\end{eqnarray}
Using the identify (\ref{eq:identity}) this can be written,
after $k^-$ integration, as
\begin{eqnarray}
1 - Z_1^{\pi\, {\rm bub}}
&=& -{1 \over 2 (4\pi f_\pi)^2}
\int\!dy\, dk_\perp^2\,
\log\left( {k_\perp^2+m_\pi^2 \over \mu^2} \right) \delta(y),
\label{eq:Z1pibuby}
\end{eqnarray}
which is equal and opposite to the pion tadpole contribution in
Eq.~(\ref{eq:Z1Ntady}).  The vanishing of the sum of the pion
tadpole and bubble contributions,
\begin{eqnarray}
(1 - Z_1^{\pi\, {\rm bub}}) + (1 - Z_1^{N\, {\rm tad}})
&=& 0,
\end{eqnarray}
ensures therefore that these have no net effect on the vertex
renormalization.  In Fig.~\ref{fig:Z1tad} the pion tadpole and
bubble diagrams are illustrated for a $k_\perp$ momentum
cutoff~$\Lambda$.

\subsection{Nonanalytic behavior}
\label{ssec:LNA}

The model-independent, nonanalytic (NA) structure of the vertex
renormalization factors can be studied by expanding $1-Z_1$ in powers
of $m_\pi$.  The terms in the expansion that are even powers of $m_\pi$
are analytic in the quark mass $m_q$ (from the Gell-Mann--Oakes--Renner
relation, $m_\pi^2 \sim m_q$ for small $m_\pi$), while odd powers of
$m_\pi$ or logarithms of $m_\pi$ are nonanalytic in $m_q$.
The NA terms reflect the long-range structure of chiral loops, and
are exactly calculable in terms of low energy constants such as $g_A$
and $f_\pi$, independent of the details of short-range physics.

The NA behavior of the nucleon rainbow contribution is given by
\begin{eqnarray}
(1-Z_1^N)
&\stackrel{\rm NA}{\longrightarrow}&
{ g_A^2 M^2 \over (4\pi f_\pi)^2 }
\left\{ {m_\pi^2 \over M^2} \log m_\pi^2\
        -\ {5\pi \over 4} {m_\pi^3 \over M^3}\
        -\ {3 m_\pi^4 \over 4 M^4} \log m_\pi^2
\hspace*{2cm}
           \left[ {1 \over D_\pi D_N^2}\ {\rm term} \right]
\right.         					\nonumber\\
& & \hspace*{1.4cm}
        -\ {m_\pi^2 \over 2 M^2} \log m_\pi^2\
        +\ {\pi \over 2} {m_\pi^3 \over M^3}\
        +\ {m_\pi^4 \over 4 M^4} \log m_\pi^2
\hspace*{2cm}
           \left[ {1 \over D_\pi D_N}\ {\rm term} \right]
                        			        \nonumber\\
& & \hspace*{1.4cm}
\left.
        +\ {m_\pi^2 \over 4 M^2} \log m_\pi^2
\right\}
\hspace*{6.4cm}
           \left[ {1 \over D_\pi}\ {\rm term} \right]
 			                               \nonumber\\
&=& { 3 g_A^2 \over 4 (4\pi f_\pi)^2 }
\left\{ m_\pi^2 \log m_\pi^2\
        -\ \pi {m_\pi^3 \over M}\
        -\ {2 m_\pi^4 \over 3 M^2} \log m_\pi^2
	+\ {\cal O}(m_\pi^5)
\right\},
\label{eq:Z1Nlna}
\end{eqnarray}
where for completeness we have included the first three lowest
order NA terms, and the origins of the various powers of $m_\pi$
have been indicated in the brackets to the right of the equations.
Comparison with the PS result (the $1/D_\pi D_N^2$ term in
Eq.~(\ref{eq:Z1Nlna})) shows that for the leading NA (LNA) term
(order $m_\pi^2 \log m_\pi^2$) one has
\begin{eqnarray}
\Big(1-Z_1^N\Big)_{\rm LNA}
&=& {3 \over 4} \Big(1-\widetilde{Z}_1^N\Big)_{\rm LNA}.
\end{eqnarray}
This makes clear the origin of the difference between the results
in Refs.~\cite{DLY70, TMS00}, which were obtained for a PS $\pi N$
coupling, and Refs.~\cite{XDJ01} and \cite{Sav02}, which were
obtained for the PV theory.
Note that this result does not depend on the details of the ultraviolet
regulator, since the nonanalytic structure is determined entirely by
the infrared behavior of the integrals.

The NA behavior of the pion rainbow contribution is given, to order
in $m_\pi$, by
\begin{eqnarray}
(1-Z_1^\pi)
&\stackrel{\rm NA}{\longrightarrow}&
{ g_A^2 M^2 \over (4\pi f_\pi)^2 }
\left\{ {m_\pi^2 \over M^2} \log m_\pi^2\
        -\ {5\pi \over 4} {m_\pi^3 \over M^3}\
        -\ {3 m_\pi^4 \over 4 M^4} \log m_\pi^2
\hspace*{2.0cm}
           \left[ {1 \over D_\pi^2 D_N}\ {\rm term} \right]
\right.                         			\nonumber\\
& & \hspace*{1.25cm}
\left.
        -\ {m_\pi^2 \over 4 M^2} \log m_\pi^2
\right\}
\hspace*{6.5cm} \left[ {1 \over D_\pi}\ {\rm term} \right]
							\nonumber\\
&=& { 3 g_A^2 \over 4 (4\pi f_\pi)^2 }
\left\{ m_\pi^2 \log m_\pi^2\
        -\ {5\pi \over 3} {m_\pi^3 \over M}\
        -\ {m_\pi^4 \over M^2} \log m_\pi^2
	+\ {\cal O}(m_\pi^5)
\right\}.
\label{eq:Z1pilna}
\end{eqnarray}
Note that the behavior arising from the $1/D_\pi^2 D_N$ term is
identical to that from the $1/D_\pi D_N^2$ term in $Z_1^N$, which
reflects the gauge invariance of the PS theory, Eq.~(\ref{eq:PSgi}).

For the PV theory, the Kroll-Ruderman terms has the nonanalytic
behavior
\begin{eqnarray}
(1-Z_1^{\rm KR})
&\stackrel{\rm NA}{\longrightarrow}&
{ g_A^2 M^2 \over (4\pi f_\pi)^2 }
\left\{ -\ {m_\pi^2 \over 2 M^2} \log m_\pi^2\
        +\ {\pi \over 2} {m_\pi^3 \over M^3}\
        +\ {m_\pi^4 \over 4 M^4} \log m_\pi^2
\hspace*{1.3cm} \left[ {1 \over D_\pi D_N}\ {\rm term} \right]
\right.                                                 \nonumber\\
& & \hspace*{1.8cm}
\left.
        +\ {m_\pi^2 \over 2 M^2} \log m_\pi^2
\right\}
\hspace*{5.7cm} \left[ {1 \over D_\pi}\ {\rm term} \right]              
							\nonumber\\
&=& { 3 g_A^2 \over 4 (4\pi f_\pi)^2 }
\left\{ {2\pi \over 3} {m_\pi^3 \over M}\
        -\ {m_\pi^4 \over 3 M^2} \log m_\pi^2
        +\ {\cal O}(m_\pi^5)
\right\}.
\label{eq:Z1KRlna}
\end{eqnarray}
Here the ${\cal O}(m_\pi^2 \log m_\pi^2)$ terms cancel between the
$1/D_\pi D_N$ and $1/D_\pi$ terms, so that the leading NA behavior
of the KR term $\sim m_\pi^3$.
For the sum of the $Z_1^\pi$ and $Z_1^{\rm KR}$ terms,
\begin{eqnarray}
(1-Z_1^\pi) + (1-Z_1^{\rm KR})
&\stackrel{\rm NA}{\longrightarrow}&
{ 3 g_A^2 \over 4 (4\pi f_\pi)^2 }
\left\{ m_\pi^2 \log m_\pi^2
	-\ \pi {m_\pi^3 \over M}\
        -\ {2 m_\pi^4 \over 3 M^2} \log m_\pi^2
\right\},
\end{eqnarray}
the NA behavior is therefore explicitly verified to be equivalent
to that in $(1-Z_1^N)$ in Eq.~(\ref{eq:Z1Nlna}).

\begin{table}[tb]
\caption{Leading nonanalytic contributions to the vertex renormalization
	$1-Z_1$, in units of $1/(4\pi f_\pi)^2\, m_\pi^2 \log m_\pi^2$. 
	The asterisks $(^*)$ in the $Z_1^N$ and $Z_1^\pi$ rows denote
	contributions that are present for the pseudoscalar $\pi N$
	coupling.
	Note that the nonanalytic contributions from the KR terms
	cancel at this order, but are nonzero at ${\cal O}(m_\pi^3)$,
	and are needed to ensure gauge invariance of the theory,
	Eq.~(\ref{eq:PVgi}).
	The sum of all contributions is given in the last two
	columns for the PV and PS theories, respectively.}
\vspace*{0.5cm}    
\begin{tabular}{l||c|c|c|c||c||c}		\hline
        &\ $1/D_\pi D_N^2$\ \
        &\ $1/D_\pi^2 D_N$\ \
        &\ $1/D_\pi D_N$\ \
        &\ $1/D_\pi$ or $1/D_\pi^2$\ \
        &\ \ sum (PV)\ \
        &\ \ sum (PS)\ \		\\ \hline
$1-Z_1^N$
        & $g_A^{2\ *}$
	& 0
	& $-\frac{1}{2} g_A^2$
	& $\frac{1}{4} g_A^2$
	&\ \ $\frac{3}{4} g_A^2$
	&\ \ $g_A^2$			\\
$1-Z_1^\pi$
        & 0
	& $g_A^{2\ *}$
	& 0
	&\!\!\!\!$-\frac{1}{4} g_A^2$
	&\ \ $\frac{3}{4} g_A^2$
	&\ \ $g_A^2$			\\
$1-Z_1^{\rm KR}$
        & 0
	& 0
	& $-\frac{1}{2} g_A^2$
	& $\frac{1}{2} g_A^2$
	&\ \ 0
	&\ \ 0				\\
$1-Z_1^{N\, {\rm tad}}$
        & 0
	& 0
	& 0
	&\!\!\!\!$-1/2$
	& $-1/2$\ \
	&\ \ 0				\\
$1-Z_1^{\pi\, {\rm bub}}$
        & 0
	& 0
	& 0
	& 1/2
	&\ \ 1/2
	&\ \ 0				\\ \hline
\end{tabular}   
\vspace*{0.5cm}
\label{tab:budget}
\end{table}

Finally, for the pion loop contributions from Sec.~\ref{ssec:Z1tad},
the NA behavior is given by the $m_\pi^2 \log m_\pi^2$ term for
both the pion tadpole (Fig.~\ref{fig:loops}(f)) and bubble
(Fig.~\ref{fig:loops}(g)) diagrams,
\begin{eqnarray}
(1-Z_1^{N\, {\rm tad}})
&\stackrel{\rm NA}{\longrightarrow}&
-{ 1 \over 2 (4\pi f_\pi)^2 }\ m_\pi^2 \log m_\pi^2,	\\
(1-Z_1^{\pi\, {\rm bub}})
&\stackrel{\rm NA}{\longrightarrow}& \hspace*{0.35cm}
{ 1 \over 2 (4\pi f_\pi)^2 }\ m_\pi^2 \log m_\pi^2.
\label{eq:Z1tadlna}
\end{eqnarray}

The results for the leading order nonanalytic contributions to the
vertex renormalization factors are summarized in Table~\ref{tab:budget}
for each of the above terms, where the entries are given in units of
$1/(4\pi f_\pi)^2\, m_\pi^2 \log m_\pi^2$.
Note that the contributions from the terms associated with
$1/D_\pi D_N^2$ and $1/D_\pi^2 D_N$ (denoted by asterisks $^*$) are
the same as those in the PS theory, while the other contributions
arise only for PV coupling.  The table clearly illustrates the origin
of the difference between the $1-Z_1^N$ corrections in the PV and PS
theories, and in particular the relative factor 3/4 found in heavy
baryon chiral perturbation theory \cite{XDJ01, Sav02} compared
with calculations based on the Sullivan process with PS coupling
\cite{Sul72, DLY70}.

\section{Parton Distributions and Moments}
\label{sec:me}

The above results on the vertex renormalization factors can be used
to compute the NA behavior of moments of parton distribution functions
(PDFs) arising from the pion cloud of the nucleon.
The presence of the pion cloud induces corrections to the PDFs of a
bare nucleon, whose Bjorken $x$ dependence can be represented in terms
of convolutions of pion and nucleon light-cone distribution functions
$f_i(y)$ and the corresponding parton distributions in the pion and
nucleon \cite{Bur12}.
The light-cone distribution $f_i(y)$ are defined such that when
integrated over $y$ they give the appropriate vertex renormalization 
factors $Z_1^i$, $(1-Z_1^i) = \int dy\, f_i(y)$.
Unlike in the PS coupling models \cite{Sul72, Tho83, Hen90, Zol91,
Mel93, Hol96, Spe97, Kum98, AM12}, the dressed nucleon PDFs in the
PV theory contain several additional terms \cite{Bur12, Moi13},
\begin{eqnarray}
q(x)
&=& Z_2\, q_0(x)\
 +\ \left( [f_N + f_{N\, \rm tad}] \otimes q_0 \right)(x)
 +\ \left( [f_\pi + f_{\pi\, \rm bub}] \otimes q_\pi \right)(x)
 +\ \left( f_{\rm KR} \otimes q_{\rm KR} \right)(x),	\nonumber\\
& &
\label{eq:conv}
\end{eqnarray}
where $q_0$ is the bare nucleon PDF (arising from the diagram in
Fig.~\ref{fig:loops}(a)),
$q_\pi$ is the PDF in the pion, and
$q_{\rm KR} \equiv \Delta q_0/g_A$, with $\Delta q_0$ the
spin-dependent PDF in the bare nucleon.
The constant $Z_2$ is the wave function renormalization constant,
and the symbol $\otimes$ represents the convolution integral
$(f \otimes q)(x) =  \int_x^1 (dz/z)\, f(z)\, q(x/z)$, where for
the $f_\pi$ and $f_{\pi\, \rm bub}$ contributions the integration
variable $z$ should be taken to be the fraction of the nucleon's
$+$ component of momentum carried by the pion, $y = k^+/p^+$,
while for the $f_N$, $f_{N\, \rm tad}$ and $f_{\rm KR}$ terms
$z = 1-y$.
Note that because of the additional pion field at the vertex
in the KR diagram, Fig.~\ref{fig:loops}(e), the contribution
of the KR terms to the (unpolarized) PDF involves a convolution
with a spin-dependent parton distribution.
In the corresponding ``Sullivan'' process based on the PS coupling
\cite{Sul72, Tho83, Hen90, Zol91, Mel93, Hol96, Spe97}, only the
$f_N$ and $f_\pi$ functions contribute, and these are related
by Eq.~(\ref{eq:PSgi}).

\subsection{Pionic corrections to twist-2 matrix elements}

According to the operator product expansion in QCD, the moments of
PDFs are related to matrix elements of local operators,
\begin{eqnarray}
\langle N | \widehat{{\cal O}}_q^{\mu_1 \cdots \mu_n} | N \rangle
&=& 2 \langle x^{n-1} \rangle_q\ p^{\{ \mu_1} \cdots p^{\mu_n \}}
\end{eqnarray}
where the braces $\{ \cdots \}$ denote symmetrization of Lorentz
indices, and the operators are given by the quark bilinears
\begin{eqnarray}
\widehat{{\cal O}}_q^{\mu_1 \cdots \mu_n}
&=& \bar\psi \gamma^{ \{ \mu_1 }\, iD^{\mu_2} \cdots iD^{\mu_n \}} \psi\
 -\ {\rm traces},
\end{eqnarray}
with $D^\mu$ the covariant derivative.
The $n$-th moment of the PDF $q(x)$ is given by
\begin{eqnarray}
\langle x^{n-1} \rangle_q
&=& \int_0^1 dx\, x^{n-1}\, (q(x) + (-1)^n \bar q(x)).
\end{eqnarray}
For $n=1$, we define $\langle x^0 \rangle_q \equiv {\cal M}^{(p)}$
to be the moment in the proton, with $u \leftrightarrow d$ for the
neutron ${\cal M}^{(n)}$.

For the direct coupling of the photon to the nucleon, which includes
the wave function renormalization (Fig.~\ref{fig:loops}(b)),
the nucleon rainbow diagram (Fig.~\ref{fig:loops}(c)), and the
pion tadpole diagram (Fig.~\ref{fig:loops}(f)),
the contributions to the twist-2 matrix elements are given by
\begin{subequations}
\begin{eqnarray}
{\cal M}^{(p)}_N
&=& Z_2 + (1-Z_1^N) + (1-Z_1^{N\, {\rm tad}}),
\label{eq:MNp}					\\
{\cal M}^{(n)}_N
&=& \hspace*{0.75cm}   2 (1-Z_1^N) - (1-Z_1^{N\, {\rm tad}}),
\label{eq:MNn}
\end{eqnarray}
\end{subequations}
for the proton and neutron, respectively.
The wave function renormalization $Z_2$ factor in Eq.~(\ref{eq:MNp}) is
\begin{eqnarray}
1-Z_2 &=& (1-Z_1^p) + (1-Z_1^n)\ =\ 3 (1-Z_1^N).
\label{eq:Z2}
\end{eqnarray}
The contributions from the photon--pion couplings, including
the pion rainbow diagram (Fig.~\ref{fig:loops}(d)),
the Kroll-Ruderman term (Fig.~\ref{fig:loops}(e)), and the
pion bubble diagram (Fig.~\ref{fig:loops}(g)), are given for
the proton and neutron by
\begin{subequations}
\begin{eqnarray}
{\cal M}^{(p)}_\pi
&=& \hspace*{0.3cm}
     2 (1-Z_1^\pi) + 2 (1-Z_1^{\rm KR}) + (1-Z_1^{\pi\, {\rm bub}}),
\label{eq:Mpip}					\\
{\cal M}^{(n)}_\pi
&=& -2 (1-Z_1^\pi) - 2 (1-Z_1^{\rm KR}) - (1-Z_1^{\pi\, {\rm bub}}).
\label{eq:Mpin}
\end{eqnarray}
\end{subequations}

Using Eq.~(\ref{eq:Z2}), the pion cloud contributions to the isoscalar
(sum of proton and neutron) moments from coupling involving nucleons
cancel,
\begin{eqnarray}
{\cal M}^{(p+n)}_N &=& 1,
\label{eq:MN_is}
\end{eqnarray}
leaving the charge of the nucleon (or valence quark number)
unrenormalized from that given by the bare coupling,
Fig.~\ref{fig:loops}(a).
Similarly, the contributions to the isoscalar moments involving direct
coupling to pions add to zero, as required by charge conservation,
\begin{eqnarray}
{\cal M}^{(p+n)}_\pi &=& 0.
\label{eq:Mpi_is}
\end{eqnarray}
Note that these results, Eqs.~(\ref{eq:MN_is}) and (\ref{eq:Mpi_is}),
are true to all orders in the pion mass, not just for the LNA parts
that were discussed in Refs.~\cite{XDJ01, Sav02, XDJer}, and to which
we turn to in the next section.

\subsection{LNA behavior of isovector moments}

The LNA contributions from the nucleon coupling diagrams to the proton
and neutron moments are given by
\begin{subequations}
\begin{eqnarray}
{\cal M}^{(p)}_N
&\stackrel{\rm LNA}{\longrightarrow}&
1 - {(3 g_A^2 + 1) \over 2 (4\pi f_\pi)^2}\, m_\pi^2 \log m_\pi^2,
\label{eq:MNpLNA}				\\
{\cal M}^{(n)}_N
&\stackrel{\rm LNA}{\longrightarrow}& \hspace*{0.75cm}
{(3 g_A^2 + 1) \over 2 (4\pi f_\pi)^2}\, m_\pi^2 \log m_\pi^2.
\label{eq:MNnLNA}
\end{eqnarray}
\end{subequations}
Taking the difference between the proton and neutron moments,
the isovector contribution then becomes
\begin{eqnarray}
{\cal M}^{(p-n)}_N
&\stackrel{\rm LNA}{\longrightarrow}&
1 - {(3 g_A^2 + 1) \over (4\pi f_\pi)^2}\, m_\pi^2 \log m_\pi^2,
\label{eq:MNlna}
\end{eqnarray}
which agrees with the results obtained in heavy baryon chiral
perturbation theory \cite{XDJ01, Sav02}.

Similarly, the LNA contributions from the pion coupling diagrams
to the proton and neutron moments are given by
\begin{subequations}
\begin{eqnarray}
{\cal M}^{(p)}_\pi
&\stackrel{\rm LNA}{\longrightarrow}& \hspace*{0.3cm}
{(3 g_A^2 + 1) \over 2 (4\pi f_\pi)^2}\, m_\pi^2 \log m_\pi^2,	\\
{\cal M}^{(n)}_\pi
&\stackrel{\rm LNA}{\longrightarrow}&
-{(3 g_A^2 + 1) \over 2 (4\pi f_\pi)^2}\, m_\pi^2 \log m_\pi^2,
\end{eqnarray}
\end{subequations}
so that the isovector contribution can be written
\begin{eqnarray}
{\cal M}^{(p-n)}_\pi
&\stackrel{\rm LNA}{\longrightarrow}&
{(3 g_A^2 + 1) \over (4\pi f_\pi)^2}\, m_\pi^2 \log m_\pi^2.
\label{eq:Mpilna}
\end{eqnarray}
The pion coupling contributions to the moment therefore cancel those
of the nucleon coupling in Eq.~(\ref{eq:MNlna}), such that the total
lowest moment of the PDF is not affected by pion loop corrections.

The analysis is more straightforward for the PS theory, where neither
tadpoles, bubbles nor KR terms are present, and LNA behavior of the
nucleon and pion coupling contributions to the isovector moments is
given by
\begin{subequations}
\begin{eqnarray}
\widetilde{{\cal M}}^{(p-n)}_N
&\stackrel{\rm LNA}{\longrightarrow}&
1 - {4 g_A^2 \over (4\pi f_\pi)^2}\, m_\pi^2 \log m_\pi^2,	\\
\label{eq:MNlnaPS}
\widetilde{{\cal M}}^{(p-n)}_\pi
&\stackrel{\rm LNA}{\longrightarrow}&
\hspace*{0.75cm}
    {4 g_A^2 \over (4\pi f_\pi)^2}\, m_\pi^2 \log m_\pi^2.
\label{eq:MpilnaPS}
\end{eqnarray}
\end{subequations}
This agrees with the results obtained in Ref.~\cite{TMS00} using
the light-cone momentum distributions computed for the Sullivan
process in the PS theory \cite{Sul72, DLY70}.
As observed in Ref.~\cite{XDJ01}, the PV and PS results agree in
the limit as $g_A \to 1$, although they clearly differ in the
general case for $g_A \neq 1$.

For higher moments, $n>1$, the pion coupling contributions will be
suppressed by additional powers of $m_\pi^2$, while the LNA behavior
of the nucleon coupling diagrams remains
$\sim m_\pi^2 \log m_\pi^2$ \cite{XDJ01, Sav02, XDJer}.
Cancellation will therefore not occur between the nucleon and pion
coupling contributions for these moments, so that the shape of
underlying PDFs will in general be modified by the presence of pion
loops.  We have verified that our LNA results agree with those
presented in Ref.~\cite{Moi13}.
Further details of the pion light-cone momentum distribution $f_i(y)$
are discussed in Ref.~\cite{Bur12}, and an analysis of their
phenomenological consequences will be presented in a forthcoming
publication \cite{CRJ13}.

\section{Conclusion}
\label{sec:conc}

In this work we have presented a detailed analysis of pion cloud
corrections to the electromagnetic coupling of the nucleon, using the
lowest order effective Lagrangian constrained by the chiral symmetry
of QCD.  We have computed the complete set of vertex corrections
arising from the various one-loop diagrams, including rainbow diagrams
with pion and nucleon coupling, Kroll-Ruderman contributions,
and tadpole and bubble diagrams associated with $\pi N$ contact
interactions.

Explicit evaluation of the vertex renormalization factors allowed us to
directly verify relations between the nucleon and pion coupling diagrams,
and demonstrate the consistency of the theory with electromagnetic gauge
invariance.  The KR terms in particular are essential for ensuring gauge
invariance to all orders in the pion mass, even though these do not
contribute to the leading nonanalytic behavior of the vertex factors.
We have also shown that the sum of the pion tadpole and bubble diagrams
vanishes.

We have examined the chiral expansion of all the vertex corrections as
a function of the pion mass $m_\pi$, computing the coefficients of the
nonanalytic terms up to and including order $m_\pi^4 \log m_\pi$.
The LNA terms agree with earlier calculations in heavy baryon chiral
perturbation theory \cite{XDJ01, Sav02}, although our formulation is
relativistic and allows for higher order corrections in $m_\pi/M$.
Comparison of the results for the pseudoscalar $\pi N$ theory reveals
the origin of the long-standing discrepancy between the LNA behavior
in the chiral effective theory and in approaches based on the
Sullivan process \cite{Sul72, DLY70, Spe97, Kum98} which use a
$\gamma_5$ coupling.

To study the behavior of the total vertex corrections, rather than
just their longest-range LNA contributions, we have computed the
vertex renormalization factors numerically as a function of the
transverse momentum cutoff used to regularize the integrals.
The pion and nucleon rainbow corrections give positive contributions to
the vertex renormalization factor $(1-Z_1)$ for the range of cutoffs
considered here ($\Lambda \leq 1$~GeV), while the contribution from
the KR diagram is negative.  The overall magnitude of the vertex
correction is
  $(1-Z_1^N) = (1-Z_1^\pi) + (1-Z_1^{\rm KR}) \approx 15\%$
for $\Lambda=0.5$~GeV and $\approx 25\%$ for $\Lambda=1$~GeV.
The tadpole and bubble contributions range up to $\approx 30\%$
for $\Lambda=1$~GeV.
Although a transverse momentum cutoff breaks the Lorentz invariance
of the $\pi N$ theory, for the purposes of the present study it is
sufficient to illustrate the relative contributions of the various
pion loop diagrams.  For a more quantitative analysis, for example,
of the corrections to the $\bar d - \bar u$ PDF difference, a
covariant regularization scheme can be used \cite{CRJ13}.

Finally, using the results for the vertex and wave function
renormalization constants we computed the pion loop corrections to
the matrix elements of twist-2 operators for the proton and neutron,
which in the operator product expansion are related to moments of
parton distribution functions.
For the lowest moment, we demonstrated explicitly that the pion
loop corrections cancel for the isoscalar combination of moments
for the nucleon and pion couplings separately.
The isovector moments, on the other hand, were found to have the
characteristic $m_\pi^2 \log m_\pi^2$ leading dependence for both
the nucleon and pion coupling diagrams (with the sum of course
canceling, as required by charge conservation).
Again, comparison of the PV and PS results for the moments enabled
us to clearly identify the source of the difference between the
coefficients of the LNA terms in the two theories.
While the PV theory is clearly preferred by considerations of
chiral symmetry, the explicit demonstration that the PS theory
can be made consistent in this context with the introduction of
a scalar $\sigma$ field remains an interesting challenge.

The results derived here can be used in the future to investigate
the nonanalytic behavior of the nucleon PDFs, particularly the
extrapolation of calculations in lattice QCD performed at unphysically
large quark masses \cite{Det01, Bal12} to the physical region.
Our findings will also pave the way for phenomenological studies,
especially the quest for a consistent interpretation of the physics
of the pion cloud at the parton level, enabling deeper studies of
the origin of the $\bar d - \bar u$ asymmetry \cite{Bur12}.
In addition, this work will also guide investigation of the very
important asymmetry between the $s$ and $\bar s$ distributions,
with its connection to the five-quark component of the nucleon
wave function, as well as the spin-flavor asymmetry
$\Delta \bar u - \Delta \bar d$ between the polarized $\bar u$
and $\bar d$ distributions.

\section*{Acknowledgements}

We thank M.~Birse, M.~Burkardt, K.~Hendricks, V.~Lyubovitskij,
M.~Polyakov, and A.~A.~Vladimirov for helpful discussions.
This work was supported by the DOE Contract No. DE-AC05-06OR23177,
under which Jefferson Science Associates, LLC operates Jefferson Lab,
DOE Contract No. DE-FG02-03ER41260, and the Australian Research Council
through the ARC Centre of Excellence for Particle Physics at the
Terascale and Grant FL0992247.

\appendix
\section{Feynman Rules}
\label{app:Feyn}

For convenience we summarize in this appendix the complete set of
Feynman rules derived from the PV Lagrangian ${\cal L}_{\gamma\pi N}$
in Eq.~(\ref{eq:LpiNg}), needed to compute the vertex renormalization
and wave function corrections from pion loops.  The conventions
throughout this work denote the nucleon momentum by $p^\mu$ and the
pion momentum by $k^\mu$, with $e$ the electric charge on the electron.
Isospin couplings that are not listed are identically zero.

\vspace*{0.5cm}

\underline{Propagators} 

Nucleon: \vspace*{-1cm}
\begin{eqnarray}
\hspace*{7cm}
{i \over \slash{\!\!\!p} - M}
\end{eqnarray}

\vspace*{-1.5cm}
\begin{figure}[h]
\hspace*{-2cm} \includegraphics[width=3cm]{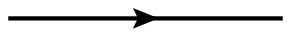}
\end{figure}

\vspace*{1cm}

Pion: \vspace*{-1cm}
\begin{eqnarray}
\hspace*{7cm}
{i \over k^2 - m_\pi^2}
\end{eqnarray}

\vspace*{-1.5cm}
\begin{figure}[h]
\hspace*{-2cm} \includegraphics[width=3cm]{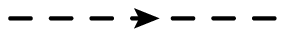}
\end{figure}

\vspace*{2cm}

\underline{$\gamma^* NN$ vertex} 

$\gamma^* p\, (n) \to p\, (n)$: \vspace*{-1cm}
\begin{eqnarray}
\hspace*{7cm}
& & -i Q_{p(n)}\, |e|\, \gamma^\mu		\\
& & {\rm with}\ \ Q_{p (n)} = 1\, (0)		\nonumber
\end{eqnarray}

\vspace*{-2.3cm}
\begin{figure}[h]
\hspace*{-2cm} \includegraphics[width=3cm]{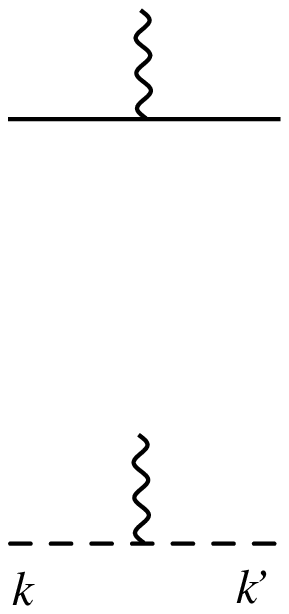}
\end{figure}

\vspace*{-3.5cm}

\underline{$\gamma^* \pi \pi$ vertex} 

$\gamma^* \pi^\pm \to \pi^\pm$: \vspace*{-1.1cm}
\begin{eqnarray}
\hspace*{7cm}
& &-i Q_{\pi^\pm}\, |e|\, (k+k')^\mu		\\
& & {\rm with}\ \ Q_{\pi^\pm}=\pm 1		\nonumber
\end{eqnarray}

\newpage
\underline{$\pi NN$ vertex} 

$p\, (n) \to \pi^0\, p\, (n)$: \vspace*{-1.1cm}
\begin{eqnarray}
\hspace*{9.5cm}
+1\, (-1)\, {g_A \over 2f_\pi} \gamma_5\, \slash{\!\!\!k}
\end{eqnarray}

\vspace*{-1.5cm}
\begin{figure}[h]
\includegraphics[width=3cm]{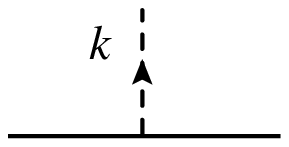}
\end{figure}
\vspace*{-0.5cm}

$p\, (n) \to \pi^+ (\pi^-)\, n\, (p)$: \vspace*{-1.1cm}
\begin{eqnarray}
\hspace*{10.7cm}
{g_A \over \sqrt{2}f_\pi} \gamma_5\, \slash{\!\!\!k}
\end{eqnarray}

\vspace*{1cm}

\underline{$\gamma^*\pi NN$ vertex} 

$\gamma^* p\, (n) \to \pi^+ (\pi^-)\, n\, (p)$: \vspace*{-1.1cm}
\begin{eqnarray}
\hspace*{10cm}
-1 (+1) {g_A |e| \over \sqrt{2}f_\pi} \gamma_5\, \gamma^\mu
\end{eqnarray}

\vspace*{-1.8cm}
\begin{figure}[h]
\includegraphics[width=3cm]{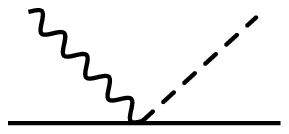}
\end{figure}

\vspace*{1cm}

\underline{$\pi\pi NN$ vertex} 

$\pi^- (\pi^+)\, p \to \pi^- (\pi^+)\, p$: \vspace*{-1cm}
\begin{eqnarray}
\hspace*{10cm}  
+1 (-1) {i \over 4 f_\pi^2} (\slash{\!\!\!k} + 
\slash{\!\!\!k'})
\label{eq:pip-pip}
\end{eqnarray}

\vspace*{-1cm}
\begin{figure}[h]
\includegraphics[width=3cm]{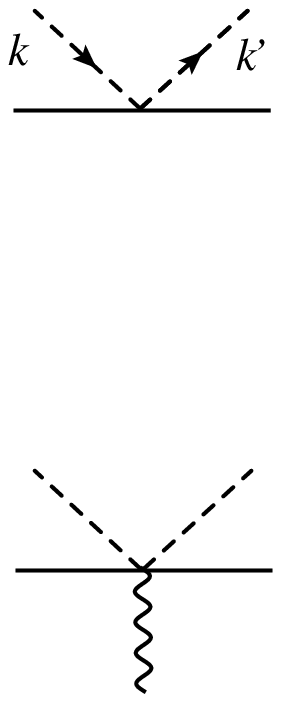}
\end{figure}

\vspace*{-7.3cm}

$\pi^- (\pi^+)\, n 
\to \pi^- (\pi^+)\, n$: \vspace*{-1cm}
\begin{eqnarray}
\hspace*{10cm}
-1 (+1) {i \over 4 f_\pi^2} (\slash{\!\!\!k} + 
\slash{\!\!\!k'})
\end{eqnarray}
\vspace*{-0.7cm}

$\pi^0\, p\, (n) \to \pi^+ (\pi^-)\, n\, (p)$: \vspace*{-1cm}
\begin{eqnarray}
\hspace*{9.5cm}
+1 (-1) {i \over 2\sqrt{2} f_\pi^2} (\slash{\!\!\!k} + 
\slash{\!\!\!k'})
\end{eqnarray}

\vspace*{1cm}

\underline{$\gamma^* \pi\pi NN$ vertex} 

$\pi^\pm\, p\, (n) \to \pi^\pm\, p\, (n)$: \vspace*{-1cm}
\begin{eqnarray}
\hspace*{10.3cm}
+1 (-1){i |e| \over 2 f_\pi^2} \gamma^\mu
\end{eqnarray}


\vspace*{0.3cm}

$\pi^0\, p\, (n) \to \pi^+ (\pi^-)\, n\, (p)$: \vspace*{-1cm}
\begin{eqnarray}
\hspace*{10.8cm}
-{i |e| \over 2 \sqrt{2} f_\pi^2} \gamma^\mu
\end{eqnarray}

\newpage
\section{Gauge Invariance}
\label{app:GI}

\begin{figure}[tb]
\includegraphics[width=10cm]{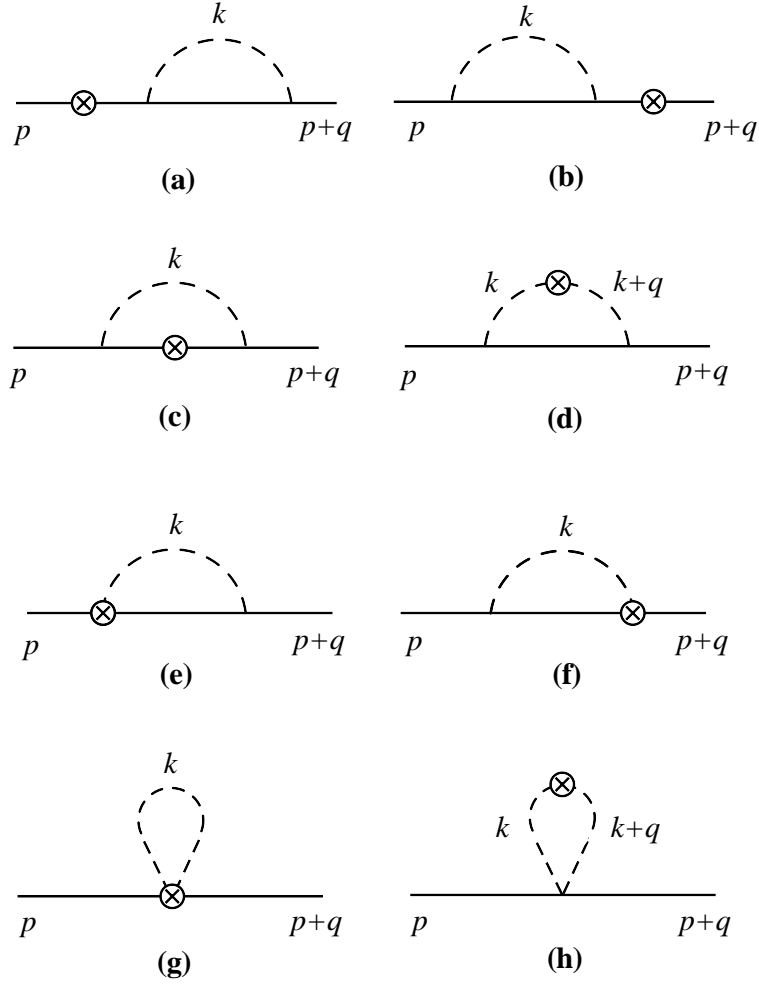}
\caption{Coupling of an electromagnetic current to a proton
	(with momentum $p$) dressed by a pion (with momentum $k$):
	(a) and (b) wave function renormalization diagrams,
	(c) rainbow diagram with coupling to the proton,
	(d) rainbow diagram with coupling to the $\pi^+$,
	(e) and (f) Kroll-Ruderman diagrams,
	(g) tadpole diagram with coupling to the $\pi\pi pp$ vertex,
	(h) bubble diagram with coupling to the pion.
	The current brings in a momentum $q$.}
\label{fig:GI}
\end{figure}

In this section we demonstrate explicitly that the electromagnetic 
coupling to the nucleon dressed by pions, illustrated in
Fig.~\ref{fig:GI}, is gauge invariant, for both the PV and PS
pion--nucleon theories.

For the $\pi^0 pp$ coupling the current derived from the PV Lagrangian
in Eq.~(\ref{eq:LpiNg}) is a sum of the three contributions in
Fig.~\ref{fig:GI}(a), (b) and (c),
\begin{eqnarray}
J^\mu_{(\pi^0)}
&=& {g_A^2 \over 4 f_\pi^2}
    \int\!\!{d^4k \over (2\pi)^4}
    \left[ {\cal J}_{\rm wf (L)}^\mu
	 + {\cal J}_{\rm wf (R)}^\mu
	 + {\cal J}_N^\mu
    \right],
\label{eq:Jpi0}
\end{eqnarray}
where
\begin{subequations}
\label{eq:Jwf} 
\begin{eqnarray}  
{\cal J}_{\rm wf (L)}^\mu
&=& \bar u(p+q)\,
    k\!\!\!\slash \gamma_5\,
    {i \over p\!\!\!\slash + q\!\!\!\slash - k\!\!\!\slash - M}\,
    \gamma_5 k\!\!\!\slash\,
    {i \over p\!\!\!\slash + q\!\!\!\slash - M}\,
    (-i e \gamma^\mu)\,
    u(p)
    {i \over D_\pi(k)},
\label{eq:JwfL}				\\
{\cal J}_{\rm wf (R)}^\mu
&=& \bar u(p+q)\,
    (-i e \gamma^\mu)\,
    {i \over p\!\!\!\slash - M}\,
    k\!\!\!\slash \gamma_5\,
    {i \over p\!\!\!\slash - k\!\!\!\slash - M}\,
    \gamma_5 k\!\!\!\slash\,
    u(p)
    {i \over D_\pi(k)},
\label{eq:JwfR}
\end{eqnarray}  
\end{subequations}
represent the ``wave function renormalization'' diagrams in
Fig.~\ref{fig:GI}(a) and (b), and
\begin{eqnarray}
\label{eq:JN}
{\cal J}_N^\mu
&=& \bar u(p+q)\,
    k\!\!\!\slash \gamma_5\,
    {i \over p\!\!\!\slash + q\!\!\!\slash - k\!\!\!\slash - M}\,
    (-i e \gamma^\mu)\,
    {i \over p\!\!\!\slash - k\!\!\!\slash - M}\,
    \gamma_5 k\!\!\!\slash\,
    u(p)
    {i \over D_\pi(k)}
\end{eqnarray}
corresponds to the rainbow diagram with coupling to the proton in
Fig.~\ref{fig:GI}(c).  Here the electromagnetic current operator
brings in a finite momentum $q$ to the proton.
Contracting the currents with the photon four-vector $q_\mu$ and
using the Dirac equation $(p\!\!\!\slash - M) u(p) = 0$, one finds
\begin{subequations}
\label{eq:qJpi0}
\begin{eqnarray}
q_\mu\, {\cal J}_{\rm wf (L)}^\mu
&=& e\, \bar u(p+q)\,
    \gamma_5 k\!\!\!\slash\,
    {1 \over p\!\!\!\slash + q\!\!\!\slash - k\!\!\!\slash - M}\,
    \gamma_5 k\!\!\!\slash\,
    u(p)
    {1 \over D_\pi(k)},
\label{eq:qJwfL}				\\
q_\mu\, {\cal J}_{\rm wf (R)}^\mu
&=& - e\, \bar u(p+q)\,
    \gamma_5 k\!\!\!\slash\,
    {1 \over p\!\!\!\slash - k\!\!\!\slash - M}\,
    \gamma_5 k\!\!\!\slash\,
    u(p)
    {1 \over D_\pi(k)},
\label{eq:qJwfR}
\end{eqnarray}
and
\begin{eqnarray}
q_\mu\, {\cal J}_N^\mu
&=& e\, \bar u(p+q)\,
    \gamma_5 k\!\!\!\slash\,
    \left(
      {1 \over p\!\!\!\slash - k\!\!\!\slash - M}
    - {1 \over p\!\!\!\slash + q\!\!\!\slash - k\!\!\!\slash - M}
    \right)
    \gamma_5 k\!\!\!\slash\,
    u(p)
    {1 \over D_\pi(k)}
\label{eq:qJN}					\\
&=& -q_\mu\, {\cal J}_{\rm wf (L)}^\mu\
 -\  q_\mu\, {\cal J}_{\rm wf (R)}^\mu,		\nonumber
\end{eqnarray}
\end{subequations}  
where we have used
$q\!\!\!\slash
= (p\!\!\!\slash + q\!\!\!\slash - M)
- (p\!\!\!\slash - M)$
and
$q\!\!\!\slash
= (p\!\!\!\slash + q\!\!\!\slash - k\!\!\!\slash - M)
- (p\!\!\!\slash - k\!\!\!\slash - M)$
to simplify Eqs.~(\ref{eq:qJwfR}) and (\ref{eq:qJN}), respectively.
The sum of the three contributions then gives the
required result,
\begin{eqnarray}
q_\mu\, J^\mu_{(\pi^0)} &=& 0.
\end{eqnarray}

For the $\pi^+ np$ coupling the current has seven contributions,
including the wave function renormalization diagrams in
Fig.~\ref{fig:GI}(a) and (b), the $\pi^+$ rainbow diagram in
Fig.~\ref{fig:GI}(d), the Kroll-Ruderman contributions in
Fig.~\ref{fig:GI}(e) and (f), and the $\pi^+$ tadpoles and bubbles in
Fig.~\ref{fig:GI}(g) and (h),
\begin{eqnarray}
J^\mu_{(\pi^+)}
&=& \int\!\!{d^4k \over (2\pi)^4}
\left\{
    {g_A^2 \over 2 f_\pi^2}\,
    \Big[ {\cal J}_{\rm wf (L)}^\mu
	+ {\cal J}_{\rm wf (R)}^\mu
	+ {\cal J}_\pi^\mu
	+ {\cal J}_{\rm KR (L)}^\mu
	+ {\cal J}_{\rm KR (R)}^\mu
    \Big]
\right.					\nonumber\\
& & \hspace*{1.33cm}
\left.
 +\ {1 \over 4 f_\pi^2}
    \Big[ {\cal J}_{N\, \rm tad}^\mu
	+ {\cal J}_{\pi\, \rm bub}^\mu
    \Big]
\right\}.
\label{eq:Jpi+tot}
\end{eqnarray}
The current for the $\pi^+$ rainbow is given by
\begin{eqnarray}
{\cal J}_\pi^\mu
&=& \bar u(p+q)\,
    (k\!\!\!\slash + q\!\!\!\slash) \gamma_5\,
    {i \over p\!\!\!\slash - k\!\!\!\slash - M}\,
    \gamma_5 k\!\!\!\slash\,
    u(p)
    {i \over D_\pi(k+q)}\,
    (-i e) (2k^\mu+q^\mu)\,
    {i \over D_\pi(k)},
\label{eq:Jpi+}
\end{eqnarray}
while the Kroll-Ruderman currents are
\begin{subequations}
\label{eq:JpiKR}
\begin{eqnarray}  
{\cal J}_{\rm KR (L)}^\mu
&=& \bar u(p+q)\,
    k\!\!\!\slash \gamma_5\,
    {i \over p\!\!\!\slash + q\!\!\!\slash - k\!\!\!\slash - M}\,
    (-e \gamma_5 \gamma^\mu)\,
    u(p)
    {i \over D_\pi(k)},
\label{eq:JKRL}					\\
{\cal J}_{\rm KR (R)}^\mu
&=& \bar u(p+q)\,
    (e \gamma_5 \gamma^\mu)\,
    {i \over p\!\!\!\slash - k\!\!\!\slash - M}\,
    \gamma_5 k\!\!\!\slash\,
    u(p)
    {i \over D_\pi(k)}.
\label{eq:JKRR}
\end{eqnarray}
\end{subequations}
Finally, for the pion tadpole and bubble diagrams the two currents
are given by
\begin{subequations}
\label{eq:Jtad}
\begin{eqnarray}
{\cal J}^\mu_{N\, \rm tad}
&=& -i\,
    \bar u(p+q)\,
    (2 k\!\!\!\slash + q\!\!\!\slash)
    u(p)
    (-i e) (2k^\mu+q^\mu)\,
    {i \over D_\pi(k)}
    {i \over D_\pi(k+q)},
\label{eq:JNtad}			\\
{\cal J}^\mu_{\pi\, \rm bub}
&=& 2i\,
    \bar u(p+q)\,
    e\, \gamma^\mu\,
    u(p)
    {i \over D_\pi(k)},
\label{eq:Jpibub}
\end{eqnarray}
\end{subequations}
respectively.
Note that diagrams involving a pion tadpole with the photon coupling
to a proton in the initial or final state directly involve loop
integrations with an odd number of factors $k$ in the integrand
(see Eq.~(\ref{eq:pip-pip})) and therefore vanish identically.

Contracting the $\pi^+$ currents with the photon momentum $q_\mu$,
one has
\begin{subequations}
\label{eq:qJpi}
\begin{eqnarray}
q_\mu\, {\cal J}_\pi^\mu
&=& e\, \bar u(p+q)\,
    \gamma_5 (k\!\!\!\slash + q\!\!\!\slash)\,
    {1 \over p\!\!\!\slash - k\!\!\!\slash - M}\,
    \gamma_5 k\!\!\!\slash\,
    u(p)
    \left( {1 \over D_\pi(k)} - {1 \over D_\pi(k+q)} \right),
\label{eq:qJpi+}                               \\
q_\mu\, {\cal J}_{\rm KR (L)}^\mu
&=& -e\, \bar u(p+q)\,
    \gamma_5 k\!\!\!\slash\,
    {1 \over p\!\!\!\slash + q\!\!\!\slash - k\!\!\!\slash - M}\,
    \gamma_5 q\!\!\!\slash\,
    u(p)
    {1 \over D_\pi(k)},
\label{eq:qJKRL}				\\
q_\mu\, {\cal J}_{\rm KR (R)}^\mu
&=& -e\, \bar u(p+q)\, 
    \gamma_5 q\!\!\!\slash\,
    {1 \over p\!\!\!\slash - k\!\!\!\slash - M}\,                
    \gamma_5 k\!\!\!\slash\,
    u(p)
    {1 \over D_\pi(k)}.
\label{eq:qJKRR}
\end{eqnarray}
\end{subequations}
Adding the wave function renormalization contributions in
Eqs.~(\ref{eq:qJwfL}) and (\ref{eq:qJwfR}), one can verify,
after some tedious but straightforward manipulations, that
\begin{eqnarray}
q_\mu\, J^\mu_{(\pi^+)} &=& 0.
\end{eqnarray}
Note that the inclusion of the KR contributions is vital to cancel
the contributions from the $\pi^+$ current with the PV coupling,
without which the theory would not be gauge invariant.
%
%
The tadpole and bubble contributions are each independently gauge
invariant, as can be seen from the contractions
\begin{subequations}
\label{eq:qJtad}
\begin{eqnarray}
q_\mu\, {\cal J}^\mu_{N\, \rm tad}
&=& e\,
    \bar u(p+q)\,
    (2 k\!\!\!\slash + q\!\!\!\slash)
    u(p)
    \left( {1 \over D_\pi(k)} - {1 \over D_\pi(k+q)} \right)\
 =\ 0,
\label{eq:qJNtad}			\\
q_\mu\, {\cal J}^\mu_{\pi\, \rm bub}
&=& -2e\,
    \bar u(p+q)\,
    q\!\!\!\slash\,
    u(p)
    {i \over D_\pi(k)}\
 =\ 0,
\label{eq:qJpibub}
\end{eqnarray}
\end{subequations}
where the first expression can be verified by changing variables
$k'\!\!\!\!\slash = k\!\!\!\slash + q\!\!\!\slash \to k\!\!\!\slash$,
and the second vanishes because of the Dirac equation.
The same of course holds true also for a neutron initial state.

For the PS coupling, only the diagrams in Fig.~\ref{fig:GI}(a)--(d)
are present.  As in the PV case, for the $\pi^0 pp$ coupling the
current in the PS theory has three contributions from
Fig.~\ref{fig:GI}(a), (b) and (c),
\begin{eqnarray}
\widetilde{J}^\mu_{(\pi^0)}
&=& g_{\pi NN}^2
    \int\!\!{d^4k \over (2\pi)^4}
    \Big[ \widetilde{{\cal J}}_{\rm wf (L)}^\mu
        + \widetilde{{\cal J}}_{\rm wf (R)}^\mu
        + \widetilde{{\cal J}}_N^\mu
    \Big],  
\label{eq:Jpi0PS}
\end{eqnarray}
where
\begin{subequations}
\label{eq:JwfPS}
\begin{eqnarray}
\widetilde{{\cal J}}_{\rm wf (L)}^\mu
&=& \bar u(p+q)\,
    i \gamma_5\,
    {i \over p\!\!\!\slash + q\!\!\!\slash - k\!\!\!\slash - M}\,
    i \gamma_5\,
    {i \over p\!\!\!\slash + q\!\!\!\slash - M}\,
    (-i e \gamma^\mu)\,
    u(p)
    {i \over D_\pi(k)},
\label{eq:JwfLPS}			\\
\widetilde{{\cal J}}_{\rm wf (R)}^\mu
&=& \bar u(p+q)\,
    (-i e \gamma^\mu)\,
    {i \over p\!\!\!\slash - M}\,
    i \gamma_5\,
    {i \over p\!\!\!\slash - k\!\!\!\slash - M}\,
    i \gamma_5\,
    u(p)
    {i \over D_\pi(k)},
\label{eq:JwfRPS}			\\
\widetilde{{\cal J}}_N^\mu
&=& \bar u(p+q)\,
    i \gamma_5\,
    {i \over p\!\!\!\slash + q\!\!\!\slash - k\!\!\!\slash - M}\,
    (-i e \gamma^\mu)\,
    {i \over p\!\!\!\slash - k\!\!\!\slash - M}\,
    i \gamma_5\,
    u(p)
    {i \over D_\pi(k)}.
\label{eq:JNPS}
\end{eqnarray}
\end{subequations}
Contracting the currents (\ref{eq:JwfPS}) with $q_\mu$ and again
using the Dirac equation, one finds
\begin{subequations}
\label{eq:qJpi0PS}
\begin{eqnarray}
q_\mu\, \widetilde{{\cal J}}_{\rm wf (L)}^\mu
&=& e\, \bar u(p+q)\,
    \gamma_5\,
    {1 \over p\!\!\!\slash + q\!\!\!\slash - k\!\!\!\slash - M}\,
    \gamma_5\,
    u(p)
    {1 \over D_\pi(k)},
\label{eq:qJwfLPS}				\\
q_\mu\, \widetilde{{\cal J}}_{\rm wf (R)}^\mu
&=& - e\, \bar u(p+q)\,
    \gamma_5\,
    {1 \over p\!\!\!\slash - k\!\!\!\slash - M}\,
    \gamma_5\,
    u(p)
    {1 \over D_\pi(k)},
\label{eq:qJwfRPS}				\\
q_\mu\, \widetilde{{\cal J}}_N^\mu
&=& e\, \bar u(p+q)\,
    \gamma_5\,
    \left(
      {1 \over p\!\!\!\slash - k\!\!\!\slash - M}
    - {1 \over p\!\!\!\slash + q\!\!\!\slash - k\!\!\!\slash - M}
    \right)
    \gamma_5\,
    u(p)
    {1 \over D_\pi(k)}.
\label{eq:qJNPS}
\end{eqnarray}
\end{subequations}  
From Eqs.~(\ref{eq:qJpi0PS}) it is evident that
\begin{eqnarray}
q_\mu\, \widetilde{{\cal J}}_N^\mu
&=& -q_\mu\, \widetilde{{\cal J}}_{\rm wf (L)}^\mu\
 -\ q_\mu\, \widetilde{{\cal J}}_{\rm wf (R)}^\mu,
\end{eqnarray}
so that gauge invariance is satisfied explicitly for the
proton dressed by a neutral $\pi^0$ with PS coupling,
\begin{eqnarray}
q_\mu\, \widetilde{J}^\mu_{(\pi^0)} &=& 0.
\end{eqnarray}

Similarly for the PS $\pi^+ np$ coupling, the current is given by
the three contributions in Fig.~\ref{fig:GI}(a), (b) and (d),
\begin{eqnarray}
\widetilde{J}^\mu_{(\pi^+)}
&=& 2 g_{\pi NN}^2
    \int\!\!{d^4k \over (2\pi)^4}
    \Big[ \widetilde{{\cal J}}_{\rm wf (L)}^\mu
        + \widetilde{{\cal J}}_{\rm wf (R)}^\mu
        + \widetilde{{\cal J}}_\pi^\mu
    \Big],
\label{eq:Jpi+totPS}
\end{eqnarray}
where
\begin{eqnarray}
\widetilde{{\cal J}}_\pi^\mu
&=& \bar u(p+q)\,
    i \gamma_5\,
    {i \over p\!\!\!\slash - k\!\!\!\slash - M}\,
    i \gamma_5\,
    u(p)
    {i \over D_\pi(k+q)}\,
    (-i e) (2k^\mu+q^\mu)\,
    {i \over D_\pi(k)}.
\label{eq:Jpi+PS}
\end{eqnarray}
The contraction of $q_\mu$ with $\widetilde{{\cal J}}_\pi^\mu$ gives
\begin{eqnarray}
q_\mu\, \widetilde{{\cal J}}_\pi^\mu
&=& e\, \bar u(p+q)\,
    \gamma_5\,
    {i \over p\!\!\!\slash - k\!\!\!\slash - M}\,
    \gamma_5\,
    \left( {1 \over D_\pi(k)} - {1 \over D_\pi(k+q)} \right),
\end{eqnarray}
where the first term in the parentheses cancels with 
$q_\mu\, \widetilde{{\cal J}}_{\rm wf (R)}^\mu$
in Eq.~(\ref{eq:qJwfRPS}), and the second cancels
with $q_\mu\, \widetilde{{\cal J}}_{\rm wf (L)}^\mu$
in (\ref{eq:qJwfLPS}) after changing variables
$k'\!\!\!\!\slash = k\!\!\!\slash + q\!\!\!\slash \to k\!\!\!\slash$.
Therefore gauge invariance is explicitly verified also
for the $\pi^+$ with PS coupling,
\begin{eqnarray}
q_\mu\, \widetilde{J}^\mu_{(\pi^+)} &=& 0.
\end{eqnarray}
Note that since these results are obtained at the operator level,
they are independent of the particular renormaliation scheme chosen
to regulate the integrals.

\section{Ward-Takahashi Identity}
\label{app:WTI}

In this section we demonstrate the consistency of the vertex corrections
and wave function renormalization with the Ward-Takahashi identity.
To begin with, we consider the nucleon self-energy operator, which can
in general be written in terms of the vector and scalar components as
\begin{eqnarray}
\hat\Sigma(p) &=& \Sigma_v\, p\!\!\!\slash\ +\ \Sigma_s.
\label{eq:Sigma_def}
\end{eqnarray}
Evaluating the matrix element of the self-energy operator between
on-shell nucleon states gives
\begin{eqnarray}
\Sigma
&=& {1\over 2}\, \sum_s \bar u(p,s)\, \hat\Sigma(p)\, u(p,s)\
 =\ M\, \Sigma_v\ +\ \Sigma_s,
\end{eqnarray}
where the sum is taken over the spin polarizations $s=\pm 1/2$ of
the nucleon.  The self-energy modifies the pole of the nucleon
propagator according to
\begin{eqnarray}
{1 \over p\!\!\!\slash - M - \hat\Sigma(p)}
&=& {Z_2 \over p\!\!\!\slash - M - \delta M},
\end{eqnarray}
where $Z_2$ is the wave function renormalization constant,
\begin{eqnarray}
Z_2 &=& {1 \over 1 - \Sigma_v},
\label{eq:Z2form}
\end{eqnarray}
and the mass shift $\delta M$ is given by
\begin{eqnarray}
\delta M &=& Z_2\, \Sigma.
\label{eq:deltaM}
\end{eqnarray}

Alternatively, using Eqs.~(\ref{eq:Z2form}) and (\ref{eq:deltaM}) one
can express the vector and scalar components of the self-energy as
\begin{eqnarray}
\Sigma_v &=& - (Z_2^{-1} - 1),
\label{eq:Sigma_v}					\\
\Sigma_s &=& (Z_2^{-1} - 1)\, M\ +\ Z_2^{-1}\, \delta M.
\label{eq:Sigma_s}
\end{eqnarray}
From Eqs.~(\ref{eq:Sigma_def}) and (\ref{eq:Sigma_v}) one observes
that the nucleon wave function renormalization $Z_2$ is related
to the derivative of the nucleon self-energy operator by
\begin{eqnarray}
(Z_2 - 1)\, \gamma^\mu
&=& { \partial \hat\Sigma(p) \over \partial p_\mu}.
\end{eqnarray}
On the other hand, from Eq.~(\ref{eq:Z1def}) in Sec.~\ref{sec:Z1}
the vertex renormalization constant $Z_1$ is defined in terms of
the matrix element of the vertex correction $\Lambda^\mu$ by
\begin{eqnarray}  
(Z_1^{-1} - 1)\, \bar u(p)\, \gamma^\mu\, u(p)
&=& \bar u(p)\, \Lambda^\mu\, u(p).
\end{eqnarray}
The Ward-Takahashi identity relates the vertex operator $\Lambda^\mu$
to the $p_\mu$ derivative of the self-energy operator $\hat\Sigma$,
which can be expressed as the equality of the vertex and wave function
renormalization factors, $Z_1 = Z_2$.

To demonstrate that this relation is explicitly satisfied by
the PV pion-nucleon theory defined by Eqs.~(\ref{eq:LpiN})
and (\ref{eq:LpiNg}), recall that the self-energy for a nucleon
dressed by a pion loop is given by \cite{JMT09}
\begin{eqnarray}
\hat\Sigma(p)
&=& 3i \left( {g_A \over 2 f_\pi} \right)^2
\int\!\!{d^4k \over (2\pi)^4}
	(\slash{\!\!\!k} \gamma_5)\,
	{i\, (\slash{\!\!\!p}-\slash{\!\!\!k}+M) \over D_N(p-k)}\,
	(\gamma_5 \slash{\!\!\!k})\,
	{i \over D_\pi(k)},
\label{eq:Sigma}
\end{eqnarray}
where $D_N(p-k) = (p-k)^2 - M^2 + i \eps$
and $D_\pi(k) = k^2 - m_\pi^2 + i\eps$.
Differentiating the nucleon propagator with respect to the nucleon
momentum, which is equivalent to the insertion of a zero energy
photon,
\begin{eqnarray}
{ \partial \over \partial p_\mu}
{ 1 \over p\!\!\!\slash - k\!\!\!\slash - M }
&=& - { 1 \over p\!\!\!\slash - k\!\!\!\slash - M }\,
      \gamma^\mu\,
      { 1 \over p\!\!\!\slash - k\!\!\!\slash - M },
\end{eqnarray}
leads to
\begin{eqnarray}
- {\partial \hat\Sigma(p) \over \partial p_\mu}
&=& 3 \left( {g_A \over 2 f_\pi} \right)^2
\int\!\!{d^4k \over (2\pi)^4}
        (\slash{\!\!\!k} \gamma_5)\,
        {i\, (\slash{\!\!\!p}-\slash{\!\!\!k}+M) \over D_N(p-k)}\,
        \gamma^\mu\, 
        {i\, (\slash{\!\!\!p}-\slash{\!\!\!k}+M) \over D_N(p-k)}\,
        (\gamma_5 \slash{\!\!\!k})\,
        {i \over D_\pi(k)}.			\nonumber\\
& &
\label{eq:dSigma_dp1}
\end{eqnarray}  
Comparing the right-hand side of (\ref{eq:dSigma_dp1}) with the
expression for $\Lambda^\mu$ in Eq.~(\ref{eq:LamN}), one can then
identify
\begin{eqnarray}
3 \Lambda^\mu_p
&=& - {\partial \hat\Sigma(p) \over \partial p_\mu}.
\label{eq:dSigma_dp2}
\end{eqnarray}
Now, changing variables $k \to p-k$ in Eq.~(\ref{eq:Sigma}) enables
the self-energy to be written equivalently as
\begin{eqnarray}
\hat\Sigma(p)
&=& 3i \left( {g_A \over 2 f_\pi} \right)^2
\int\!\!{d^4k \over (2\pi)^4}
        (\slash{\!\!\!p}-\slash{\!\!\!k}) \gamma_5\,
        {i\, (\slash{\!\!\!k}+M) \over D_N(k)}\,
        \gamma_5 (\slash{\!\!\!p}-\slash{\!\!\!k})\,
        {i \over D_\pi(p-k)},
\label{eq:Sigma_pi}
\end{eqnarray}
which when differentiated with respect to $p_\mu$ gives rise to
three terms,
\begin{subequations}
\begin{eqnarray}
- {\partial \hat\Sigma(p) \over \partial p_\mu}
&=& 3 \left( {g_A \over 2 f_\pi} \right)^2
\int\!\!{d^4k \over (2\pi)^4}
\left\{ 2 (p-k)^\mu (\slash{\!\!\!p}-\slash{\!\!\!k}) \gamma_5\,
	{ i\, (\slash{\!\!\!k}+M) \over D_N(k) }\,
	\gamma_5 (\slash{\!\!\!p}-\slash{\!\!\!k})\,
	\left( {i \over D_\pi(p-k)} \right)^2
\right.					\nonumber\\
& & \hspace*{3cm}
     -\ i\, \gamma^\mu\, \gamma_5\,
	{i\, (\slash{\!\!\!k}+M) \over D_N(k)}\,
        \gamma_5 (\slash{\!\!\!p}-\slash{\!\!\!k})\,
        {i \over D_\pi(p-k)}		\nonumber\\
& & \hspace*{3cm}
\left.
     -\ (\slash{\!\!\!p}-\slash{\!\!\!k})\, \gamma_5\,
	{i\, (\slash{\!\!\!k}+M) \over D_N(k)}\,
	i\, \gamma_5\, \gamma^\mu\,
	{i \over D_\pi(p-k)}
\right\}.
\label{eq:dSigma_dp3}
\end{eqnarray}
Changing variables $k \to p-k$ once again then gives
\begin{eqnarray}  
3\, {1 \over 2}
    \left( \Lambda_{\pi^+}^\mu\ +\ \Lambda_{{\rm KR}, \pi^+}^\mu
    \right)
&=& - {\partial \hat\Sigma(p) \over \partial p_\mu}.
\label{eq:dSigma_dp4}
\end{eqnarray}
\end{subequations}
Since the expressions for $\hat\Sigma(p)$ in Eqs.~(\ref{eq:Sigma})
and (\ref{eq:Sigma_pi}) are equivalent, this implies that the
electromagnetic operators are related by
\begin{eqnarray}
\Lambda_p^\mu
&=& {1 \over 2}
    \left( \Lambda_{\pi^+}^\mu\ +\ \Lambda_{{\rm KR}, \pi^+}^\mu       
    \right).
\label{eq:Lam_WTI}
\end{eqnarray}
Taking matrix elements of both sides of Eq.~(\ref{eq:Lam_WTI})
between proton states, one arrives at the relation
\begin{eqnarray}
(1-Z_1^p) &=& {1\over 2} (1-Z_1^{\pi^+})
	   +  {1\over 2} (1-Z_1^{{\rm KR}, \pi^+}),
\label{eq:WTI}
\end{eqnarray}
or, in terms of the ``isoscalar'' vertex factors defined
in Sec.~\ref{sec:Z1}, the desired expression
$(1-Z_1^N) = (1-Z_1^\pi) + (1-Z_1^{\rm KR})$
as in Eq.~(\ref{eq:PVgi}).

The proof for the PS case follows similarly, and is in fact more
straightforward since the KR terms are absent in this case.
One can also check the validity of the Ward-Takahashi identity
by performing the integrations explicitly, and comparing the
integrated expressions for the vertex renormalization factors.

\section{Nonanalytic Behavior of Integrals}
\label{app:lna}

In this appendix we summarize some useful results for the integrals
appearing in the expressions for the vertex corrections in
Sec.~\ref{sec:Z1} and the self-energies Appendix~\ref{app:WTI},
using dimensional regularization to render ultraviolet divergent
integrals finite.
Expanding the results in powers of $m_\pi$, we also provide explicit
expressions for their nonanalytic behavior in the chiral limit.

For the integrals involving only the pion propagator ($1/D_\pi$),
\begin{eqnarray}
\int d^4k\ {1 \over D_\pi}
&=& i \pi^2 m_\pi^2
\left[ {1 \over \eps} + 1 - \gamma - \log\pi
     - \log{m_\pi^2 \over \mu^2} + {\cal O}(\eps)
\right]						\nonumber\\
&\stackrel{\rm NA}{\longrightarrow}&
-i \pi^2 m_\pi^2 \log m_\pi^2,			
\label{eq:Dpi}					\\
& &						\nonumber\\
\int d^4k\ {1 \over D_\pi^2}
&=& i \pi^2
\left[ {1 \over \eps} - \gamma - \log\pi
     - \log{m_\pi^2 \over \mu^2} + {\cal O}(\eps)
\right]						\nonumber\\
&\stackrel{\rm NA}{\longrightarrow}&        
-i \pi^2 \log m_\pi^2.
\label{eq:Dpi2}
\end{eqnarray}
The expressions in Eqs.~(\ref{eq:Dpi}) and (\ref{eq:Dpi2}) can also
be related using
\begin{eqnarray}
{\partial \over \partial m_\pi^2} \int d^4k\ {1 \over D_\pi}
&=& \int d^4k\ {1 \over D_\pi^2}.
\end{eqnarray}
Note that there are no higher order contributions in $m_\pi$
for either of these integrals.
The corresponding integrals involving only the nucleon propagator
($1/D_N$), on the other hand, do not have nonanalytic contributions,
\begin{eqnarray}
\int d^4k\ {1 \over D_N}
&=& i \pi^2 M^2
\left[ {1 \over \eps} + 1 - \gamma - \log\pi
     - \log{M^2 \over \mu^2} + {\cal O}(\eps)
\right]						\nonumber\\
&\stackrel{\rm NA}{\longrightarrow}& 0,		\\
& &						\nonumber\\
\int d^4k\ {1 \over D_N^2}
&=& i \pi^2
\left[ {1 \over \eps} - \gamma - \log\pi
     - \log{M^2 \over \mu^2} + {\cal O}(\eps)
\right]						\nonumber\\
&\stackrel{\rm NA}{\longrightarrow}& 0.
\end{eqnarray}
For the integral of one pion propagator and one nucleon propagator,
one has
\begin{eqnarray}
\int d^4k\ {1 \over D_\pi D_N}
&=& i \pi^2
\left[ {1 \over \eps} + 2 - \gamma - \log\pi - \log{M^2 \over \mu^2}
     - {m_\pi^2 \over 2M^2} \log{m_\pi^2 \over M^2}
\right.						\nonumber\\
& & \hspace*{1cm}
\left.
     - {m_\pi\, r \over M^2}
       \left( \tan^{-1}{m_\pi \over r}
	    + \tan^{-1}{2M^2-m_\pi^2 \over m_\pi r}
       \right)
\right]						\nonumber\\
&\stackrel{\rm NA}{\longrightarrow}&
-{i \pi^2 \over 2M^2}
  \left[ m_\pi^2 \log m_\pi^2 + 2 \pi M m_\pi + {\cal O}(m_\pi^3)
  \right],
\label{eq:DpiDN}
\end{eqnarray}
where $r = \sqrt{4M^2-m_\pi^2}$.
Note that this term is responsible for the leading ${\cal O}(m_\pi^3)$
behavior of the nucleon self-energy \cite{JMT09}.

Finally, for integrals involving one pion and two nucleon propagators,
or one nucleon and two pion propagators, the additional powers of the
loop momentum render the results finite,
\begin{eqnarray}
\int d^4k\ {1 \over D_\pi^2 D_N}
&=& -i \pi^2 {1 \over M^2}
\left[ {1 \over 2} \log{m_\pi^2 \over M^2}
     - {m_\pi^2 - 2M^2 \over m_\pi r}
       \left( \tan^{-1}{m_\pi \over r}
	    + \tan^{-1}{2M^2-m_\pi^2 \over m_\pi r}
       \right)
\right]						\nonumber\\
&\stackrel{\rm NA}{\longrightarrow}&
- {i \pi^2 \over 2M^2}
  \left[ \log m_\pi^2 + {\pi M \over m_\pi} + {\cal O}(m_\pi)
  \right],
\label{eq:Dpi2DN}				\\
& &						\nonumber\\
\int d^4k\ {1 \over D_\pi D_N^2}
&=& i \pi^2 {1 \over M^2}
\left[ {1 \over 2} \log{m_\pi^2 \over M^2}
     - {m_\pi \over r}
       \left( \tan^{-1}{m_\pi \over r}
            + \tan^{-1}{2M^2-m_\pi^2 \over m_\pi r}
       \right)
\right]						\nonumber\\
&\stackrel{\rm NA}{\longrightarrow}&
{i \pi^2 \over 2M^2}
  \left[ \log m_\pi^2 - {\pi m_\pi \over 2 M} + {\cal O}(m_\pi^3)
  \right].
\label{eq:DpiDn2}
\end{eqnarray}
The expressions in Eqs.~(\ref{eq:DpiDN}) and (\ref{eq:Dpi2DN})
can also be related using
\begin{eqnarray}
{\partial \over \partial m_\pi^2} \int d^4k\ {1 \over D_\pi D_N}
&=& \int d^4k\ {1 \over D_\pi^2 D_N}.
\end{eqnarray}


\end{document}